\documentclass[a4paper,12pt,english]{article}
\usepackage[T1]{fontenc}
\usepackage[latin1]{inputenc}

\usepackage{babel}

\providecommand{\LyX}{L\kern-.1667em\lower.25em\hbox{Y}\kern-.125emX\@}

\usepackage{amssymb,amsthm}
\usepackage{exscale}
\usepackage{latexsym}
\newtheorem{Def}{Definition}                            
\newtheorem{Cl}{Claim}
\newtheorem{Prop}{Proposition}
\newtheorem{Conclusion}{Conclusion}
\newtheorem{Theorem}{Theorem}
\newtheorem{Lemma}{Lemma}
\newtheorem{Remark}{Remark}
\newcommand{\bDef}{\begin{Def}}
\newcommand{\eDef}{\end{Def}}
\newcommand{\bcl}{\begin{Cl}}
\newcommand{\ecl}{\end{Cl}}
\newcommand{\bprop}{\begin{Prop}}
\newcommand{\eprop}{\end{Prop}}
\newcommand{\bpr}{\begin{proof}}
\newcommand{\epr}{\end{proof}}
\newcommand{\bconcl}{\begin{Conclusion}}
\newcommand{\econcl}{\end{Conclusion}}
\newcommand{\btheorem}{\begin{Theorem}}
\newcommand{\etheorem}{\end{Theorem}}
\newcommand{\bla}{\begin{Lemma}}
\newcommand{\ela}{\end{Lemma}}
\newcommand{\brm}{\begin{Remark}}
\newcommand{\erm}{\end{Remark}}
\newcommand{\be}{\begin{equation}}
\newcommand{\ee}{\end{equation}}

\newcommand{\tril}{\triangleleft}
\newcommand{\ar}{\longrightarrow}
\newcommand{\bear}{\begin{array}}
\newcommand{\ear}{\end{array}}
\newcommand{\ten}{\otimes}

\begin{document}

\newcommand{\rt}{\triangleright }

\newcommand{\lt}{\triangleleft }
 
\newcommand{\te}{\otimes }

\newcommand{\rp}{\rightarrow }

\newcommand{\ra}{\mapsto }

%%%%%%%%%%%%%%%%%%%%  Harold's conventions   %%%%%

\def\zetbar{\bar{z}}
\def\z{\zeta}
\def\zbar{{\bar{\zeta}}}
\def\zetbar{{\bar{z}}}
\def\oz{\overline{z}}
\def\rep{representation }

\newcommand{\eq}[1]{(\ref{#1})}
\def\nn{\nonumber}
\def\bea{\begin{eqnarray}}
\def\eea{\end{eqnarray}}
\def\beqa{\begin{eqnarray}} 
\def\eeqa{\end{eqnarray}} 
\def\beq{\begin{equation}} 
\def\eeq{\end{equation}} 
\def\bit{\begin{itemize}}
\def\eit{\end{itemize}}
\def\obar{\overline}
\def\tens{\otimes}

\newtheorem{prop}{Proposition}[section]
\newtheorem{theorem}[prop]{Theorem}
\newtheorem{lemma}[prop]{Lemma}

\def\R{{\mathbb R}}
\def\C{{\mathbb C}}
\def\N{{\mathbb N}}
\def\Z{{\mathbb Z}}
\def\one{\mbox{1 \kern-.59em {\rm l}}}

\def\a{\alpha}          \def\da{{\dot\alpha}}
\def\b{\beta}           \def\db{{\dot\beta}}
\def\d{\delta}  \def\D{\Delta}  \def\ddt{\dot\delta}
\def\e{\epsilon}                \def\vare{\varepsilon}
\def\f{\phi}    \def\F{\Phi}    \def\vvf{\f}
\def\h{\eta}
\def\k{\kappa}
\def\l{\lambda} \def\L{\Lambda} \def\la{\lambda}
\def\m{\mu}     \def\n{\nu}
\def\o{\omega}
\def\p{\pi}     \def\P{\Pi}
\def\r{\rho}
\def\s{\sigma}  \def\S{\Sigma}
\def\t{\tau}
\def\th{\theta} \def\Th{\Theta} \def\vth{\vartheta}
\def\X{\Xeta}
\def\z{\zeta}

\thispagestyle{empty}

\rightline{LMU-TPW 2003-09}
\rightline{MPP-2003-68}

\vspace{4em}
\begin{center}

{\Large{\textbf{Gauge Field Theory on the \( E_{q}(2)- \)covariant Plane}}}

\vskip 4em

{{\textbf{Frank Meyer${}^{a,b,}$\footnote{meyerf@theorie.physik.uni-muenchen.de}, Harold
Steinacker${}^{a,}$\footnote{hsteinac@theorie.physik.uni-muenchen.de}} }}

\vskip 2em

${}^{a}$Universit\"at M\"unchen, Fakult\"at f\"ur Physik\\
        Theresienstr.\ 37, D-80333 M\"unchen, Germany\\[1em]

${}^{b}$Max-Planck-Institut f\"ur Physik\\
        F\"ohringer Ring 6, D-80805 M\"unchen, Germany\\[1em]

\end{center}

\vspace{4em}

\begin{abstract}

Gauge theory on the \( q- \)deformed two-dimensional Euclidean plane
$\R^2_q$ is studied using two different approaches. We first
formulate the theory using the natural algebraic
structures on $\R^2_q$, such as a covariant differential calculus,
a frame of one-forms and invariant integration.
We then consider a suitable star product, and introduce a
natural way to implement the Seiberg-Witten map. 
In both approaches, gauge invariance requires a suitable ``measure''
in the action, breaking the $E_{q}(2)$-invariance.
Some possibilities to avoid this conclusion using additional terms in
the action are proposed. 

\end{abstract}

\newpage
\setcounter{page}{1}
\section{Introduction}

Gauge theories provide the best known descriptions of the fundamental
forces in nature. At very short distances however, physics is not known, 
and it is plausible that spacetime is 
quantized below some scale. This idea has been contemplated 
for quite some time, and %received a boost recently
gauge theory on noncommutative spaces has been the 
subject of much research activity, see e.g.
\cite{Douglas:2001ba} for a review. 

There are several different approaches to 
gauge theories on noncommutative spaces: First, one can
formulate the theory in terms of the algebraic structures which define
the noncommutative space, such as the noncommutative algebra of functions,
its modules, and differential calculi.
Gauge transformations can then be defined by unitary elements of 
the algebra of functions. 
Examples of noncommutative gauge theories using 
this formulation can be found in 
\cite{Krajewski:1999ja,Connes:1998cr,Douglas:2001ba}. 
While it is certainly very natural, this approach
seems to be restricted to unitary gauge groups, and 
the set of admissible representations of the associated matter fields is
also quite restricted. 

Another approach has been developed following the discovery that 
string theory leads to noncommutative gauge theories 
under suitable conditions, as explained in \cite{Seiberg:1999vs}. This 
lead to a technique expressing the noncommutative fields in terms 
of commutative ones, and writing the  Lagrangians in terms of 
ordinary (commutative) fields and star products. 
It allows to formulate models with general gauge groups 
and representations, including the standard model \cite{Calmet:2001na}.
However, the Lagrangians become increasingly complicated at each order in the 
deformation parameter, and there is generally a large amount of arbitrariness
in these actions.
Moreover, the formulation of gauge theories on
general noncommutative spaces with non-constant Poisson structure
is less clear. 
In particular, no satisfactory formulation of gauge theory
on spaces with  quantum group symmetry has been given; see  e.g. 
\cite{Pawelczyk:2003nb} for a clear manifestation of this problem.
It seems that in general, a satisfactory implementation of
generalized symmetries (quantum group symmetries) in noncommutative 
field theory is yet to be found.

In the present paper, we apply these different approaches 
to gauge theory on the Euclidean quantum plane $\R^2_q$, which 
is covariant under the $q$-deformed two-dimensional Euclidean 
group $E_q(2)$. This is one of the simplest quantum spaces with 
a non-trivial quantum group symmetry, and scalar field theory on $\R^2_q$
has already been studied in \cite{Chaichian:1999wy}. 
It seems therefore well suited to gain some insights into gauge theory
on spaces with quantum group symmetry.

We first try to formulate (abelian) gauge  theory on $\R^2_q$
using an algebraic approach, taking advantage of the covariant differential
calculus on $\R^2_q$. This leads very naturally to a definition
of gauge fields and their field strength, with gauge transformations
being the unitaries of the algebra of functions. This field strength
reduces to the usual one in the commutative limit. However, the 
definition of an invariant action turns out to be less clear: if one uses the 
natural invariant integral on $\R^2_q$, one must add a 
nontrivial ``measure function'' in order to obtain a gauge invariant
action. This measure function 
explicitly breaks translation invariance, which seems to be a 
generic feature of gauge theory on spaces with quantum group symmetry.
Hence gauge invariance appears to be in conflict with quantum group 
symmetry. However, we point out some ways to avoid this conclusion. 
We propose a model with an additional scalar (``Higgs'') field
with a suitable potential,
which is manifestly gauge invariant and restores the 
formal $E_q(2)$-invariance while spontaneously breaking gauge
invariance. 

In the second part of this paper, 
we apply the star product approach to gauge theory on  $\R^2_q$, 
expressing all fields in terms of commutative ones. 
We first construct a suitable star product, and study its properties and
the relation with the integral. The gauge theory is then formulated
using this star product
in close analogy to the algebraic approach. In particular, 
the noncommutative calculus suggests a definition of the field 
strength in terms of a ``frame'', which 
ensure the correct classical limit. This is somewhat different from 
other approaches proposed in the literature \cite{Schraml:2002fi}. 
The corresponding Seiberg-Witten maps are solved up to first order.
The formulation of a gauge invariant action requires again a
nontrivial measure function, which is essentially the same as in the 
algebraic approach. While it cannot be canceled as in the algebraic approach
by introducing a Higgs field, we show how the 
action can be modified in order to obtain the correct commutative limit.

\section{The \protect\( q-\protect \)deformed two-dimensional
  Euclidean 
Group and Plane}

\subsection{The dual symmetry algebras $E_{q}(2)$ and $U_q(e(2))$}

We start by reviewing the quantum group \( E_{q}(2) \), 
which is a deformation of the 
(Hopf) algebra of functions on the two-dimensional Euclidean 
Group $E(2)$. It is generated by the 
``functions'' $ n,v,\bar{n},\bar{v}$ with the following relations and
structure maps \cite{Schupp:1992ex}
\begin{equation} 
\label{Eq} \begin{array}{cccc}
\qquad v\bar {v}=\bar  {v} v=1 \qquad n\bar {n}=\bar {n} n \qquad vn=qnv \\
\qquad n\bar {v}=q\bar {v} n \qquad v\bar {n}=q\bar {n} v \qquad \bar {n}\bar {v}=q\bar {v}\bar {n} \\
\Delta(n)=n\otimes\bar {v}+v\otimes n \qquad \Delta(v)=v\otimes v \qquad \Delta(\bar {n})=\bar {n}\otimes v+\bar {v}\otimes \bar {n} \\
\Delta(\bar {v})=\bar {v}\otimes \bar {v} \qquad \varepsilon(n)=\varepsilon(\bar {n})=0 \qquad \varepsilon(v)=\varepsilon(\bar {v})=1 \\
S(n)=-q^{-1}n \quad S(v)=\bar {v} \quad S(\bar {n})=-q\bar {n} \quad S(\bar {v})=v\\
\end{array} 
\end{equation}
where \( q\in \mathbb {R} \). This is a star-Hopf algebra with the conjugation
\beq
n^* = \bar{n},\qquad  v^* = \bar{v}.
\eeq
In terms of the operators \( \theta,t,\bar{t} \)
defined by \cite{Schupp:1992ex} 
\begin{equation}
v=e^{\frac{i}{2}\theta}\qquad t=nv \qquad
\bar {t}=\bar {v} \bar {n} 
\end{equation}
(note that \( v \) is unitary and can therefore be parametrized by
a hermitian element $\theta^* = \theta$),
the coproduct of \( t \) and \( \bar{t} \)
reads 
\begin{equation} \label{copro t} \Delta(t)=  t \otimes 1+ e^{i\theta} \otimes t
\qquad \Delta(\bar {t})=\bar {t} \otimes 1 + e^{-i\theta} \otimes \bar {t}.
\end{equation}
It is often convenient to consider also the dual quantum 
group. The dual Hopf algebra $U_{q}(e(2))$ of $E_{q}(2)$ is
generated by \( T,\overline{T},J \) with the following commutation
relations and structure maps\footnote{Our
  generators are related to the generators $\mu ,\nu ,\xi$ 
in \cite{Schupp:1992ex}  by
$\mu \equiv  T,  -q^{2}\nu  \equiv  \overline{T},  \xi  \equiv J$}
\cite{Schupp:1992ex}
\[
\begin{array}{ccccccccccc}
T\overline{T} & = & q^{2}\overline{T}T &  & [J,T] & = & iT &  & [J,\overline{T}] & = & -i\overline{T}
\end{array}\]
\begin{equation}
\label{eq: commrel and structure maps for Uq(e(2))}
\begin{array}{ccccccc}
\Delta (T) & = & T\te q^{2iJ}+1\te T &  & \Delta (\overline{T}) & = & \overline{T}\te q^{2iJ}+1\te \overline{T}
\end{array}
\end{equation}
\[
\begin{array}{ccccccccccc}
\Delta (J) & = & J\te 1+1\te J &  & \varepsilon (T) & = & \varepsilon (\overline{T}) & = & \varepsilon (J) & = & 0
\end{array}\]
\[
\begin{array}{ccccccccccc}
S(T) & = & -Tq^{-2iJ} &  & S(\overline{T}) & = & -\overline{T}q^{-2iJ} &  & S(J) & = & -J,
\end{array}
\]
where the dual pairing on the generators is given by\begin{equation}
\label{eq: dual pairing Uq(e(2)) and Eq(2)}
\begin{array}{cccc}
\langle T,\theta ^{i}t^{j}\overline{t}^{k}\rangle =\delta _{0i}\delta _{1j}\delta _{0k}, & \langle \overline{T},\theta ^{i}t^{j}\overline{t}^{k}\rangle =-q^{2}\delta _{0i}\delta _{0j}\delta _{1k}, & \langle J,\theta ^{i}t^{j}\overline{t}^{k}\rangle =\delta _{1i}\delta _{0j}\delta _{0k} .
\end{array}
\end{equation}
%Moreover, we have 
%\begin{equation}
%\label{eq: complex conjugation of J}
%\overline{J}=-J\, \, \, .
%\end{equation}
This is again a star-Hopf algebra with the conjugation
\[
J^* = -J, \quad T^* = \overline{T}.
%X^*=\overline{X} 
\]
%for $X\in\{ T,\overline{T},J\}$.

\subsection{The $E_{q}(2)$-covariant Euclidean plane $\R^2_q$.}

Hopf algebras can be used to define generalized symmetries.
There are two equivalent, dual notions. 
A Hopf algebra \( \mathcal{H} \) coacts on an algebra \( \mathcal{A} \)
if $\mathcal{A}$ is a left (or right) \( \mathcal{H} \)-comodule
algebra (see Appendix \ref{appendix: mathematical}) via a left coaction
$\rho: \mathcal{A} \to \mathcal{H}  \tens \mathcal{A}$.
In particular,
every Hopf algebra \( \mathcal{H} \) admits a comodule structure
on itself in virtue of the comultiplication 
\be
\Delta : \mathcal{H} \longrightarrow \mathcal{H} \otimes \mathcal{H}. 
\ee 
Observing that the 
subalgebra of $E_q(2)$ generated by \( t,\bar{t} \) is a
$E_q(2)$-module subalgebra, we can obtain the 
\( E_{q}(2)- \)symmetric plane by renaming $t \to z,\; \bar{t}
\to\bar{z}$. 
Hence $\mathbb {R}_{q}^{2}$ is the $E_{q}(2)$-comodule algebra with
generators $z,\bar{z}$ and commutation relations
\beq
\label{com z}
z\overline{z}=q^{2}\overline{z}z.
\eeq
We will also allow formal power series, and define the 
algebra of functions on the
\emph{\( E_{q}(2)- \)covariant plane} \cite{Chaichian:1999wy}
\begin{equation}
\label{de:Cext}
\mathbb {R}_{q}^{2}:=\mathbb {R} \langle \langle z,\bar{z} \rangle \rangle /(z\bar{z}-q^{2}\bar{z}z).
\end{equation}
By construction, it is covariant under 
the following left \( E_{q}(2) \)-coaction
\begin{equation}
\label{coac}
\begin{array}{ccc}
\rho (z) & = & e^{i\theta }\te z+t\te 1\\
\rho (\overline{z}) & = & e^{-i\theta }\te \overline{z}+\bar{t}\te 1.
\end{array}
\end{equation}
More formally, we have a coaction $\rho: \R^2_q \to  E_q(2) \tens \R^2_q $.
{From} now on, functions are considered to be elements of this algebra.

In general, a left comodule algebra $\mathcal{A}$ under $\mathcal{H}$
is also a right $\mathcal{H}'$-module algebra, using the dual pairing
between $\mathcal{H}$ and its dual $\mathcal{H}'$. Explicitly,
the right\footnote{Similarly one gets a left action via a dual
  pairing from a right coaction.}  action
$\tril :\mathcal{A}\ten \mathcal{H}'\ar \mathcal{A}$ 
of \( \mathcal{H}' \) on \( \mathcal{A} \) is given by
\begin{equation}
\label{de: action from coaction}
f\lt X:=(\langle X,\cdot \rangle \te id)\circ \rho (f)=\langle
X,f_{(-1)}\rangle f_{(0)}\, ,\, \, \, \, \, \, \, \, \, X\in
\mathcal{H}',\, f\in \mathcal{A}\; .
\end{equation}
Applied to the present situation using
the coaction  (\ref{coac}) and
the dual pairing (\ref{eq: dual pairing Uq(e(2)) and Eq(2)}),
we obtain an action of \( U_{q}(e(2)) \) on \( \mathbb {R}^{2}_{q} \). 
It is compatible with the conjugation $z^* = \bar{z}$ in the sense
\beq
(f\lt X)^* = f^* \lt S^{-1}(X^*)
\eeq
for any $f \in \R^2_q$ and $X \in U_q(e(2))$.
To calculate the action 
of \( U_{q}(e(2)) \) on formal power series in 
\( z,\overline{z} \), it is useful to 
note that any formal power series \( f(z,\bar{z}) \) can be
written as
\begin{equation}
\label{f-decomp}
f(z,\overline{z})=\sum _{m\in \mathbb {Z}}z^{m}f_{m}(z\overline{z}).
\end{equation}
The action on terms of this form is calculated in Appendix 
\ref{appendix: mathematical}:
\begin{eqnarray}
z^{k}f(z\overline{z})\lt T & = & \frac{z^{k-1}}{1-q^{-2}}(f(q^{2}z\overline{z})-q^{-2k}f(z\overline{z}))\nonumber \\
z^{k}f(z\overline{z})\lt \overline{T} & = & \frac{q^{4}}{1-q^{2}}z^{k+1}\frac{f(z\overline{z})-f(q^{-2}z\overline{z})}{z\overline{z}}\label{eq: action of J,T,Tbar on z^kf(zzbar)} \\
z^{k}f(z\overline{z})\lt J & = & i^{k}z^{k}f(z\overline{z})\,,\nonumber 
\label{action of Uq(e(2))}
\end{eqnarray}
which has again the above form.

\subsection{Covariant differential calculus on $\R^2_q$}
\label{sec:calculus}

A differential calculus is useful to write down  Lagrangians.
A covariant differential calculus over $\R^2_q$ is a 
graded bimodule $\Omega^*_{q} = \oplus_n \; \Omega^n_{q}$
over $\R^2_q$ which is a $U_q(e(2))$--module algebra, 
together with an exterior derivative $d$ which satisfies $d^2=0$ and 
the usual graded Leibniz rule.
Its construction \cite{Wess:1991vh,Chaichian:1994ft} 
is reviewed here for convenience,
in order to establish the notation.
We start by introducing variables \( dz \) and \( d\overline{z} \),
which are the \( q- \)differentials of \( z \) and \( \overline{z}. \) These
are noncommutative differentials which do not commute with the space
coordinates \( z,\overline{z} \). 
Covariance and $d(1) =0$ implies the coaction
\beq
\begin{array}{ccc}
\rho (dz) & = & e^{i\theta }\te dz\\
\rho (d\overline{z}) & = & e^{-i\theta }\te d\overline{z}\,,
\end{array}
\eeq
and the commutation relations between
coordinates and their differentials must be
\begin{equation}
\label{eq: comm-rel dz,dzbar with z,zbar}
\begin{array}{cccccccc}
zdz & = & q^{-2}dzz &  &  & \overline{z}dz & = & q^{-2}dz\overline{z}\\
zd\overline{z} & = & q^{2}d\overline{z}z &  &  & \overline{z}d\overline{z} & = & q^{2}d\overline{z}\, \overline{z}\,.
\end{array}
\end{equation}
To see that \( d: \mathbb {R}_{q}^{2}\rp \Omega _{q}^{1}\) 
is well-defined, we have to verify that
it respects the commutation relations of the algebra, i.e. 
\begin{equation}
\label{eq: differential d well-defined}
d(z\overline{z}-q^{2}\overline{z}z)\stackrel{!}{=}0\,
\end{equation}
which is easy to see.
To obtain a higher order differential calculus, we 
apply \( d \) on the commutation
relations (\ref{eq: comm-rel dz,dzbar with z,zbar}), which gives
\begin{equation}
\label{eq: comm-rel for differentials}
dzd\overline{z}=-q^{2}d\overline{z}dz
\end{equation}
and\[
(dz)^{2}=(d\overline{z})^{2}=0\, .
\]
This defines a star-calculus (i.e. with a reality structure), 
where the star of forms and derivatives is defined in the obvious way.
One can now introduce \emph{\( q- \)deformed partial
derivatives} by 
\begin{equation}
\label{de: exterior differential gives derivatives}
d=:dz^{i}\partial _{i}=dz\partial_z + d\bar{z}\partial_{\bar{z}}\,,
\end{equation}
as in the commutative case. This defines the action of  
\( \partial _{z} \)
and \( \partial _{\overline{z}} \)
on functions. One can also introduce
the algebra of {\em differential operators} with generators
 \( \partial _{z},\partial _{\overline{z}}, z, \bar{z} \). In order to
distinguish the generators $\partial _{z},\partial _{\overline{z}}$
in this algebra from their action on a function, we denote the
latter by 
\[
\partial _{z}(f)\, \, \, \, \, \, \textrm{and}\, \, \, \, \, \,
\partial _{\overline{z}}(f)\, ,
\]
whereas we will not 
use brackets if \( \partial _{z},\partial _{\overline{z}} \) are interpreted
as part of the algebra of differential operators.

The derivatives $\partial _{z},\partial _{\overline{z}}$
satisfy a modified Leibniz rule. It can be derived from the
Leibniz rule of the exterior differential together with the commutation
relations of differentials and coordinates as follows:
On the one hand, we have
\begin{eqnarray}
d(fg) & = & (df)g+f(dg)\nonumber \\
 & = & dz^{i}\partial _{i}(f)g+fdz^{i}\partial _{i}(g)\nonumber \\
 & = & dz^{i}\partial _{i}(f)g+dzf(q^{-2}z,q^{-2}\overline{z})\partial _{z}(g)+d\overline{z}f(q^{2}z,q^{2}\overline{z})\partial _{\overline{z}}(g)\label{eq: calculating the q-Leibniz rule} 
\end{eqnarray}
using the commutation relations
\beqa
\begin{array}{ccc}
f(z,\overline{z})\;dz & = & dz\;f(q^{-2}z,q^{-2}\overline{z})\\
f(z,\overline{z})\;d\overline{z} &=& d\overline{z}\;f(q^{2}z,q^{2}\overline{z})\,,
\end{array}
\eeqa 
which follow from (\ref{eq: comm-rel dz,dzbar with z,zbar}).
On the other hand, we have
\[
d(fg)=dz\partial _{z}(fg)+d\overline{z}\partial _{\overline{z}}(fg)\, ,
\]
and together with (\ref{eq: calculating the q-Leibniz rule})
we obtain the  {\em \( q- \)Leibniz rule}
\beqa
\label{eq: q-Leibniz rule}
\partial _{z}(fg) & = & \partial _{z}(f)g+f(q^{-2}z,q^{-2}\overline{z})\partial _{z}(g)\\
\partial _{\overline{z}}(fg) & = & \partial _{\overline{z}}(f)g+f(q^{2}z,q^{2}\overline{z})\partial _{\overline{z}}(g)\, .
\eeqa
Applying this to
the functions \( zf \) resp. \( \overline{z}f \), one
obtains the following commutation relations:\begin{equation}
\label{eq: comm rel derivatives with coordinates}
\begin{array}{cccccccc}
\partial _{z}z & = & 1+q^{-2}z\partial _{z} &  &  & \partial _{z}\overline{z} & = & q^{-2}\overline{z}\partial _{z}\\
\partial _{\overline{z}}z & = & q^{2}z\partial _{\overline{z}} &  &  & \partial _{\overline{z}}\overline{z} & = & 1+q^{2}\overline{z}\partial _{\overline{z}}\,.
\end{array}
\end{equation}
Furthermore, applying \( \partial _{z}\partial _{\overline{z}} \)
on the function \( z\overline{z} \)  we find
\begin{equation}
\label{eq: comm rel for q-derivatives}
\partial _{z}\partial _{\overline{z}}=q^{2}\partial _{\overline{z}}\partial _{z}\, .
\end{equation}
For completeness we also give the commutation relations for
differentials and derivatives: 
\begin{equation}
\label{eq: comm rel for derivatives and differentials}
\begin{array}{cccccccc}
\partial _{z}dz & = & q^{2}dz\partial _{z} &  &  & \partial _{z}d\overline{z} & = & q^{-2}d\overline{z}\partial _{z}\\
\partial _{\overline{z}}dz & = & q^{2}dz\partial _{\overline{z}} &  &  & \partial _{\overline{z}}d\overline{z} & = & q^{-2}d\overline{z}\partial _{\overline{z}}\, .
\end{array}
\end{equation}
Clearly, the \( q- \)differentials and \( q- \)derivatives become the classical
differentials resp. derivatives in the limit
\( q\rp 1 \).

\subsubsection{The frame}

On many noncommutative spaces \cite{Madore:Buch,Cerchiai:2000qu}, 
there exists
a particularly convenient basis of one--forms (a ``frame'') 
$\theta^a \in \Omega^1$, which commute with all functions.
%We are hence looking for a new basis 
%\( \theta =:\theta^{z},\overline{\theta }=:\theta ^{\overline{z}} \)
%of \( \Omega _{q}^{1} \) commuting with all functions. 
They are easy
to find here: consider the elements
\beqa
\theta \equiv \theta^{z} &:=& z^{-1}\overline{z}dz \nonumber\\
\overline{\theta } \equiv \theta^{\overline{z}} &:=& 
d\overline{z}z\overline{z}^{-1} \,.
\label{de: Frame theta^i}
\eeqa
Then the following holds:

\bla\label{la: frame}
\begin{equation}
\label{eq: theta commutes with all functions f}
[\theta ,f]=[\overline{\theta },f]=0
\end{equation}
 for all functions \( f\in \mathbb {R}_{q}^{2} \), and
\beq
\theta \overline{\theta }=-q^{2}\overline{\theta }\theta \,.
\label{theta-thetabar}
\eeq
\ela

\begin{proof}
Easy verification using the above commutation relations.
\end{proof}

It is even possible to find a one-form \( \Theta  \) which generates
the exterior differential:
consider the following ``duals'' of the frame,
\beqa
\lambda _{z} &:=&\frac{1}{1-q^{-2}}\overline{z}^{-1}\,\\
\lambda _{\overline{z}} &:=& -\frac{1}{1-q^{-2}}z^{-1}
\eeqa
and define
\[
\Theta :=\theta ^{i}\lambda _{i}\, .
\]
Then we have
\begin{Lemma}\label{la: Theta generates d}
The anti-hermitian one-form $\Theta^* = -\Theta$ generates the
exterior differential by
\beq
df=[\Theta ,f]=[\lambda _{i},f]\theta ^{i}
\label{Thetacomm}
\eeq
 for all  \( f\in \mathbb {R}_{q}^{2} \).
Similarly,
\beq
\label{Thetacomm-1}
d\alpha =\{\Theta ,\alpha \}
\eeq
for any one-form $\a$. Here \( \{\cdot ,\cdot \} \) denotes
the anti-commutator.
Furthermore,
\beq
d\Theta =\Theta ^{2}=0\, .
\label{dTheta}
\eeq
\end{Lemma}

\begin{proof}

Equations \eq{Thetacomm} and \eq{dTheta}
are shown in Appendix \ref{app:proof-La2}.
Equation \eq{Thetacomm-1} then follows easily noting that
 \( \{\Theta ,\alpha f\}=\{\Theta ,\alpha \}f-\alpha [\Theta ,f] \)
and \( \{\Theta ,f\alpha \}=[\Theta ,f]\alpha +f\{\Theta ,\alpha \}\) 
for arbitrary functions \( f \) and one-forms \( \alpha  \).

\end{proof} \textbf{}

\subsection{Invariant metric}
 
A relation between the algebra, the differential calculus and the geometry  
on noncommutative spaces was proposed in \cite{Madore:Buch}. 
We briefly address this issue here, arguing that
$\R^2_q$ is flat. This can be seen as follows:
According to \cite{Madore:Buch}, ``local'' line elements must have the form
\beq
ds^2 = \theta^i \tens \theta^j\; g_{ij}
\eeq
where $g_{ij}$ must be a central (i.e. numerical, here) 
tensor, and $\theta^i$ is the frame
introduced above. The symmetry of $g_{ij}$ is expressed in the equation
\beq
g_{ij}\; {P^{(-)}}^{ij}_{kl} =0,
\eeq
where ${P^{(-)}}^{ij}_{kl}$ is the antisymmetrizer defined by the calculus
\beq
\theta^k \theta^l\; {P^{(-)}}^{ij}_{kl} = \theta^i \theta^j.
\eeq
If we require furthermore that $ds^2$ be invariant under $E_q(2)$,
it follows that
\beq
ds^2 =  \theta \tens \overline{\theta} \, + q^2\; \overline{\theta}\tens\theta
  = q^{-2} dz \tens d\overline{z}+ q^4 d\overline{z}\tens  dz \, .
\eeq
This is certainly a flat metric, and 
for $q\rightarrow 1$ reduces to the usual Euclidean metric on $\R^2$.

\subsection{Representations of $\R^2_q$}

In the following we will only need representations of the algebra
$\R^2_q$, not including  derivatives or forms. They are easy to find \cite{Chaichian:1999wy}:
Since $r^2 = z \oz$ is formally hermitian, we assume that it can be
diagonalized.
%$r^2 \mid n\rangle  =  r^2_n \mid  n\rangle$ for some set of labels $n$. 
The commutation relations then imply 
that $z$ and $\oz$ are rising  resp. lowering operators
which are invertible,
\beqa
r^2 \mid n\rangle_{r_0}  &=&  r_0^2\; q^{2n} \mid  n\rangle_{r_0}, \nn\\
\oz \mid n\rangle_{r_0} &=& r_0  q^n \mid n+1 \rangle_{r_0}, \nn\\
z \mid n\rangle_{r_0} &=& r_0 q^{n-1} \mid n-1 \rangle_{r_0}.
\label{basic-rep}
\eeqa
We will denote this irreducible \rep  with  $L_{r_0}$,
where $r_0$ can be either positive or negative. 
The representations with $r_0$ and $-r_0$ are equivalent. The irreducible
representations are labeled by $r_0 \in [1,q)$. 
It follows that $z^{-1}$ and $\oz^{-1}$ are well-defined on $L_{r_0}$
unless $r_0=0$.

\section{Invariant Integration}

\subsection{Integral of functions}
\label{se: integral of fct}
In order to define an invariant action, 
we need an integral on $\R^2_q$
which is invariant under \( E_{q}(2) \).
In general, an integral 
(i.e. a linear functional) is called invariant with respect
to the right action of \( U_{q}(e(2)) \) if it satisfies the following
invariance condition
\begin{equation}
\label{eq: invariance condition for the q-integral}
\int ^{q}f(z,\overline{z})\lt X=\varepsilon (X)\int ^{q}f(z,\overline{z})
\end{equation}
for all \( f \in \R^2_q\) and \( X\in U_{q}(e(2)) \). 
Here $\varepsilon (X)$ is the counit.
Such an integral was found in \cite{Koe}; however, we want to  
determine the most general invariant integral here.
Since \( \varepsilon  \) is an algebra homomorphism,
it is sufficient to check the condition 
(\ref{eq: invariance condition for the q-integral})
for the generators 
\( T,\overline{T} \) and \( J \). Let us first consider functions of
the type  
\be
z^{m}f(z\bar {z}) 
\ee
where $f(r^2), r^2 = z\bar {z}$ 
can be considered as a classical function in one variable.
We can choose it such that the integral will be well-defined.
Invariance under the action \eq{action of Uq(e(2))}
of \( J \) implies 
\begin{equation}
\label{pr: q-integral for m not equal 0}
\int^{q}z^{m}f(z\overline{z}) = \d_{m,0}\;  \langle f(r^2) \rangle_r
\end{equation}
where 
$\langle f(r^2)\rangle_r$ is a ``radial'' integral to be determined. 
Invariance under the action of \( T \) and \( \overline{T} \)
then leads to the following algebraic condition 
\begin{equation}
\label{pr: q-integral}
\langle f(q^{2}r^2)-q^{-2}f(r^2) \rangle_r =0
\end{equation}
on the radial part of the integral. 
This condition is satisfied for
\begin{equation}
\label{de: Eq(2) invariant q-integral 1}
\langle f(r^2) \rangle_{r_0}:=r^{2}_{0}(q^{2}-1)\sum ^{\infty
}_{k=-\infty }q^{2k}f(q^{2k}r_{0}^{2}),
\end{equation}
for any $r_0 \in \R$.
Notice that the integral 
can then be written as ``quantum trace'' (or Jackson-sum) 
over the irreducible representation 
$L_{r_0}$ defined in \eq{basic-rep}:
\beq
\int^{q,(r_0)} f(z,\overline{z}) := (q^2-1)\; \textrm{Tr}_{r_0}(r^2f(z,\overline{z})),
\eeq
where $\textrm{Tr}_{r_0}$ is the ordinary trace on $L_{r_0}$; note that
$\textrm{Tr}_{r_0}(z^m f(r^2)) =0$ for $m \neq 0$.
If we allow superpositions of this basic integral 
(resp. direct sums of irreps of $\R^2_q$), then we can take an arbitrary
superposition of the form
\beq
\langle f(r^2) \rangle_r = \int_1^q dr_0 \mu(r_0)\; \langle f(r^2)\rangle_{r_0}
\label{jackson_int}
\eeq
with arbitrary (positive) "weight" function $\mu(r) >0$. 
If $\mu(r)$ is a delta-function,
this is simply the above Jackson-sum. For 
$\mu(r_0) = \frac 1{r_{0}(q^{2}-1)}$, 
one obtains the classical radial integral
\beq
\int^{q} f(z,\overline{z}) = \int_1^q dr_0 \frac
1{r_{0}(q^{2}-1)}\int^{q,(r_0)}f_0(z\overline{z}) =
\int_0^{\infty} dr r f_0(r^2) 
\label{class_int}
\eeq
for $f(z,\overline{z}) = \sum_m z^m f_m(r^2)$,
assuming $q>1$.
Any of these integrals reduces to the 
usual (Riemann) integral on $\R^2$ for $q\to 1$, using the obvious 
mapping from $\R^2_q$ to $\R^2$ induced by \eq{f-decomp}.

It is quite remarkable that the classical radial 
integral is indeed invariant, cp. \cite{Steinacker:1995jh}. 
This will be useful in the star-product approach
in Section \ref{Star Product Approach}.
Nevertheless, the invariant integrals are not cyclic in the ordinary sense:

\begin{Lemma}\label{la: cyclic property of q-integral}

For any invariant integral 
\eq{eq: invariance condition for the q-integral} the following 
cyclic property holds:
\bit
\item [i)] For any functions \( f,g \), we have
%\end{lyxlist}
\begin{equation}
\label{eq: cyclic property of the q-integral}
\int ^{q}fg=\int ^{q}g\mathcal{D}(f)\, 
\end{equation}
where \( \mathcal{D} \) is the algebra homomorphism defined by 
\begin{equation}
\label{de: operator D (cyclic property of the q-integral)}
\mathcal{D}(z^{m}):=q^{-2m}z^{m}\, \, \, \, \, \, \, \, \, \, \mathcal{D}(\overline{z}^{m}):=q^{2m}\overline{z}^{m}.
\end{equation}

%\begin{lyxlist}{00.00.0000}
\item [ii)] \( \mathcal{D} \) is an inner automorphism:
%can be obtained by conjugation with \( z\overline{z} \): 
\begin{equation}
\label{eq: operator D by conjugation}
\mathcal{D}(f(z,\overline{z}))=z\overline{z}f(z,\overline{z})\overline{z}^{-1}z^{-1}\, .
\end{equation}
\eit
%\end{lyxlist}
\end{Lemma}

\begin{proof}

Easy verification using 
%The equations (\ref{de: operator D (cyclic property of the q-integral)})
%and (\ref{eq: operator D by conjugation}) follow by calculation using
the commutation relation (\ref{com z}).

\end{proof}

A similar cyclic property for invariant integrals on a 
\( SO_{q}(N)- \)covariant space
was found in \cite{Steinacker:1995jh}.

\subsection{Integral of forms}

Since any 2-form
\( \alpha ^{(2)}\in \Omega ^{2}_{q} \) can be written as
\( \alpha ^{(2)}=f\theta \overline{\theta } \)
and $\theta \overline{\theta }$ is invariant, we define
\begin{equation}
\label{de: integration of forms}
\int^q \alpha ^{(2)}=\int^q f\theta \overline{\theta}:=\int ^{q}f\, .
\end{equation}
 For one-forms \( \alpha ,\beta  \) we then obtain the following cyclic
property:
\begin{equation}
\label{eq: cyclic property of forms}
\int^{q}\alpha \beta =-\int^{q}\beta \mathcal{D}(\alpha )\,
\end{equation}
where $\mathcal{D}$ is defined on forms as above. Noting that
$\mathcal{D}(\Theta) =\Theta$, this immediately yields Stokes theorem:

\begin{Theorem}

Let \( \alpha  \) be a one-form. Then \begin{equation}
\label{eq: Stokes theorem}
\int ^{q}d\alpha =0\,  .
\end{equation}

\end{Theorem}

\begin{proof}

Since \( d\alpha =\{\Theta ,\alpha \} \) 
due to \eq{Thetacomm-1},
we get with (\ref{eq: cyclic property of forms}) \[
\int ^{q}d\alpha =\int ^{q}\{\Theta ,\alpha \}=0.\]
 \end{proof}

\section{Gauge Transformations, Field Strength and Action}
\label{sec:gauge-trafos}

We consider 
matter fields as functions in \( \mathbb {R}_{q}^{2} \). An infinitesimal
noncommutative gauge transformation of a matter field \( {\psi } \)
is defined as \cite{Madore:2000en}
\begin{equation}
\label{de:trafopsi}
\delta \psi =i\Lambda \psi 
\end{equation}
while of course
$\delta z^{i}=0$.
We introduce the {}``covariant derivative'' (or rather a covariant one-form)
\begin{equation}
\label{de:cov.cord}
D:=\Theta -iA\,
\end{equation}
which should be an anti-hermitian one-form.
Requiring that \( D\psi (x) \) transforms covariantly, i.e.
\[
\delta D\psi =i\Lambda D\psi 
\]
leads to 
\beq
\delta D = i[\Lambda, D],
\label{D-gaugetrafo}
\eeq
which using \eq{Thetacomm}
implies the following gauge transformation property for the gauge
field \( A \) \begin{equation}
\label{de: trafo von A}
\delta A=[\Theta ,\Lambda ]+i[\Lambda ,A]=d\Lambda +i[\Lambda ,A]\,  .
\end{equation}
This suggests to 
define the noncommutative field strength 
\( F \)
as
\[
F:=iD^{2}  =F_{ij}\theta ^{i}\theta ^{j} \; ,
\]
which is a 2-form transforming as
\beq
\delta_{\Lambda }F= i [\Lambda, F].
\eeq
Since \( \Theta ^{2}=0 \) and \( \{\Theta ,A\}=dA \) we obtain
the familiar form
\beq
F=dA-iA^{2}\, ,
\label{F-form}
\eeq
which shows that $F$ reduces to the classical field strength 
in the limit $q \to 1$.
To write it in terms of components, it is most natural to expand
the 1-forms in the frame basis $\theta^i = (\theta, \bar{\theta})$, 
because then no ordering prescription is needed. Hence we can write 
\begin{equation}
A = A_i \theta^i  = \theta^i  A_i ,
\end{equation}
and the field strength is
\beqa
F &=& (\lambda _{i}A_{j} + A_{i}\lambda_{j} -iA_{i}A_{j})\theta ^{i}\theta ^{j} \nn\\
  &=& (\la_1 A_2 - q^{-2} A_2 \la_1 
  - q^{-2} \la_2  A_1 + A_1 \la_2 
  -i A_1 A_2 +i q^{-2}  A_2  A_1) \theta \bar{\theta}
\eeqa
where $\la_i = (\la_z, \la_\zetbar)$.
Notice that this is written in terms of the components 
of the frame, not of the differentials $dz, d \zetbar$. 
In order to understand its classical limit, it is better to 
%In the limit \( q\rp 1 \) it is better to 
write\footnote{This is 
not natural for $q\neq1$, since then $dz, d \zetbar$ do 
not commute with functions.}
\beq
A = \tilde{A}_z dz + \tilde{A}_\zetbar d\zetbar,
\eeq 
and we recover from \eq{F-form} the classical field strength
\begin{equation}
\label{eq: limit q to 1 for F equals B to the two}
F\stackrel{q\rp 1}{\longrightarrow }(\partial _{z}\tilde{A}_{\overline{z}}-\partial _{\overline{z}}\tilde{A}_{z})dzd\overline{z}\, .
\end{equation}
In order to write down a Lagrangian for a Yang-Mills theory,  we also need
the \emph{Hodge dual} \( *_{H}F \) of \( F \).
This is easy to find: since 
any two-form \( F \) can be written as 
\[
F=f\theta \overline{\theta }=q^{-2}fdzd\overline{z}\]
for some  function \( f \), we define \( *_{H} \)
on two-forms as\begin{equation}
\label{de: Hodge dual of functions}
*_{H}F:=\frac{1}{2}f\, .
\end{equation}
This is the correct definition because
\( dzd\overline{z} \) is invariant under
\( U_{q}(e(2)) \) transformations, 
hence the Hodge dual satisfies 
\begin{equation}
\label{eq: Hodge dual invariance property}
(*_{H}F)\triangleleft u=*_{H}(F\triangleleft u)
\end{equation}
for all \( u\in U_{q}(e(2)) \).
We can now write down the following action using one of the invariant integrals found in Section \ref{se: integral of fct}:
\begin{equation}
S:=\int ^{q}F(*_{H}F)\;\overline{z}^{-1}z^{-1}=\int
^{q}\frac{1}{2}f^{2}\;\overline{z}^{-1}z^{-1}\theta \overline{\theta }\,.
\label{action-algebra}
\end{equation}
The factor \( \overline{z}^{-1}z^{-1} \) is required by 
gauge invariance under \eq{de: trafo von A},
using the property 
\begin{equation}
\label{cyclic-fix}
\int ^{q}fg\;\overline{z}^{-1}z^{-1}=\int ^{q}gf\;\overline{z}^{-1}z^{-1}
\end{equation}
which follows from Lemma \ref{la: cyclic property of q-integral}.
In the classical limit we obtain 
\[
S\stackrel{q\rp 1}{\longrightarrow }\int \frac{1}{2}(\partial
_{z}A_{\overline{z}}-\partial
_{\overline{z}}A_{z})^{2}z^{-1}\overline{z}^{-1}dzd\overline{z}\, .
\]
The ``measure factor'' $\overline{z}^{-1}z^{-1}$
breaks the $E_q(2)$-invariance explicitly. Unfortunately, it is 
required by gauge invariance. 
In other words, the invariant integral seems incompatible 
with this kind of gauge invariance, and one is faced with the choice of 
giving up either gauge invariance or $E_q(2)$-invariance\footnote{In
  the classical limit, the measure function can be
written as
$\overline{z}^{-1}z^{-1}dzd\bar{z}=\frac{1}{r^2}(rdrd\varphi)=d(\mathrm{ln}r)d\varphi$,
which is the volume-form on a cylinder. Therefore this action 
could be interpreted as Yang-Mills action on a quantum cylinder. However
this is not the aim of this paper.}. In this
paper, we will insist on gauge invariance. 

There are several possibilities how this problem might be avoided. 
One may try to modify the gauge transformation, e.g. by using some kind of 
$q$-deformed gauge invariance as in \cite{Grosse:2000Fuzzy1}. 
Unfortunately we were not able to find a satisfactory prescription
here \cite{Meyer}.
Alternatively, we will propose in the next section 
a mechanism using spontaneous symmetry breaking, which 
yields an $E_q(2)$-invariant action for low energies.
In any case, the above action is certainly appealing because
of its simplicity, and the gauge transformations \eq{D-gaugetrafo} are
very natural.
This problem may also be a hint that the quantum group
spacetime-symmetry has not been correctly implemented in the field
theory, beyond a formal level. 
A proper treatment 
would presumably require a second quantization, such that the 
$E_q(2)$-symmetry acts on a many-particle Hilbert space and the quantum fields,
as in \cite{Grosse:2001pr}. 

Let us briefly discuss the critical points of the above action. 
The absolute minima are 
given by solutions of the zero curvature condition \( F=0. \)
In terms of the coordinates \( D =  D_{i} \theta^i \) this leads to 
\[
D_{2}D_{1}=q^{2}D_{1}D_{2}\,  .
\]
 This is the defining relation of the deformed Euclidean plane with
\emph{opposite} multiplication. One solution
is of course \( D=\Theta  \), and we get all possible solutions in terms
of the automorphisms of \( \mathbb {R}_{q}^{2} \).

\section{Restoring $E_q(2)$-invariance
through spontaneous symmetry breaking} \label{symmetry-breaking}

The explicit ``weight'' factor 
$\overline{z}^{-1}z^{-1}$ in \eq{action-algebra} is rather unwelcome,
because it
explicitly breaks the $E_q(2)$-invariance of the action, which was 
the starting point for our considerations.
One could in principle interpret it as some kind of 
additional ``metric'' term in the action, which is required by 
gauge invariance. 
However, it is also possible to cancel it by the 
vacuum-expectation value (VEV) of a suitable scalar field: 
Consider the action 
\beq
S_1:=\int ^{q}F(*_{H}F) e^\phi\; \overline{z}^{-1} z^{-1} \, .
\eeq
This is gauge invariant if $\phi$ transforms in the adjoint:
\beq
\phi \to i [\Lambda,\phi].
\eeq
We can then add an action for $\phi$, such as
\beq
S_2 = \int ^{q} V(\phi)\; \overline{z}^{-1} z^{-1}
\eeq
where $V(x)$ is an ordinary function, 
which is again gauge invariant.
If we could find a potential $V(\phi)$ 
which has $e^\phi = z\overline{z} $ as solution, 
we would obtain the following ``low-energy'' action
\beq
S_1[A,\langle \phi \rangle] = \int ^{q}F(*_{H}F) 
\label{on-shell}
\eeq
replacing $\phi$ by its VEV $\langle \phi \rangle$. 
This is formally invariant under 
$E_q(2)$, while the gauge invariance is spontaneously broken
rather than explicitly.
To find such a potential $V$, consider the equation of motion 
\beq
\delta S_2[\phi] = \int^q \delta\phi \, V'(\phi)\; \overline{z}^{-1} z^{-1} =0
\eeq
using the cyclic property of the integral,
where $V'$ denotes the ordinary derivative of the power series $V(x)$. 
We therefore need a potential 
$V(x)$ such that $V'(\ln(z\overline{z})) =0$. For a given irrep $L_{r_0}$
labeled by $r_0$ as in (\ref{basic-rep}), the eigenvalues of $z\overline{z}$ are 
$r^2_0 q^{2n} = e^{2n \ln(q) + 2\ln(r_0)}$ for $n \in \Z$. 
Therefore
\beq
V_{r_0}'(2n \ln(q) + 2\ln(r_0)) =0, \qquad n \in \Z.
\eeq
This certainly holds for 
$V_{r_0}'(x) \propto \sin(2\pi\frac{x-2\ln(r_0)}{2\ln(q)})$, thus
\beq
V_{r_0}(x) = -V_0\; \cos(2\pi\frac{x-2\ln(r_0)}{2\ln(q)})
\eeq
is a possible potential.
Hence we will use the representation $L_{r_0}$, and the 
quantum trace $\int^{q,(r_0)}$
on $L_{r_0}$ as invariant integral for the action.
Note furthermore that 
\beq
\delta_\phi S_1 = 0
\eeq
for $F=0$, therefore $e^\phi = z\overline{z}, F=0$ 
is indeed a possible ``vacuum'' of the combined action 
\beq
S = S_1 + S_2 = \int^{q,(r_0)} \left(F(*_{H}F) e^\phi\; + V(\phi) \right) 
      \overline{z}^{-1} z^{-1}.
\eeq
Replacing $\phi \to \langle \phi \rangle = \ln(z\overline{z})$, it
reduces to
\beq
\int ^{q,(r_0)}F(*_{H}F)\; +\; const,
\eeq
as desired. The fluctuations in $\phi$ are suppressed if $V_0$ is
chosen large enough.
Of course there are other solutions for $\phi$, which would 
give a nontrivial ``effective metric'' 
$e^{\langle\phi\rangle} \overline{z}^{-1} z^{-1}$ in the 
action. This is somewhat reminiscent of the low-energy effective actions
in string theory, where the dilaton enters in a similar way.

For reducible representations of $\R^2_q$ one could still find such 
potentials, but if we take continuous superpositions as in \eq{class_int}
in order to have the classical integral (as in the Seiberg-Witten
approach below),
this is no longer possible.

\section{Star Product Approach}
\label{Star Product Approach}

We now want to study gauge theory on $\R^2_q$ using the star product
approach, which was developed in \cite{Bayen:1978ha,Madore:2000en}.
We will denote classical variables on $\R^2$ 
by greek letters $\z,\overline{\z}$ in this
section, in order to distinguish them from the generators $z,\overline{z}$
of the algebra $\R^2_q$.

A star product corresponding to $\R^2_q$
is defined as the pull-back of the product in $\R^2_q$
via an invertible map 
\beq
\rho: \R [[\z,\obar{\z}]][[h]]\rightarrow \R^2_q
\label{rho}
\eeq
of vector spaces,
\beq
f\star g:=\rho^{-1}(\rho(f)\rho(g)),
\eeq
where 
\beq
q = e^h.
\eeq
For example, 
the star product corresponding to normal ordering in $\R^2_q$
(i.e. commuting all \( z \) to
the left and all \( \overline{z} \) to the right) 
reads \cite{Madore:2000en}
\begin{equation}
\label{de: normal odered star product}
f\star _{n}g=\mu \circ e^{-2h(\overline{\z}\partial _{\overline{\z}}\te 
\z\partial_{\z})}(f\te g)\, .
\end{equation}
For our purpose the following star product will be more useful
\begin{equation}
\label{de: q-symm star product}
f\star _{q}g:=\mu \circ e^{h(\z\partial_{\z}\te 
\zbar\partial_{\overline{\z}}-\overline{\z}\partial_{\overline{\z}}\te \z\partial _{\z})}(f\te
g)=fg+h\z \overline{\z}(\partial_{\z} f \partial_{\overline{\z}} g - \partial_{\overline{\z}} f \partial_{\z} g) +
\mathcal{O}(h^2) \, ,
\end{equation}
because it is  hermitian, i.e. 
\( \overline{f\star _{q}g}=\overline{g}\star _{q}\overline{f} \),
and satisfies other nice properties 
as shown in Lemma \ref{la: measure for trace property} (see equation
(\ref{eq: measure for integral}) below).
The corresponding Poisson structure reads\footnote{The Poisson  structure 
is given by 
\( [f\stackrel{\star _{q}}{,}g]=ih\theta ^{ij}\partial _{i}f\partial _{j}g+O(h^{2}) \).}
\begin{equation}
\label{eq: Poisson structure}
\theta ^{ij}=-2i\z\overline{\z}\varepsilon ^{ij}\, .
\end{equation}
This star product is equivalent to the normal 
ordered one (\ref{de: normal odered star product})
via the equivalence transformation
\[
T:=e^{-h\z\partial _{\z}\overline{\z}\partial _{\overline{\z}}}\, .
\]
To see this, we first note that 
\[
\z\partial _{\z}\overline{\z}\partial _{\overline{\z}}\circ \mu =\mu
\circ (\z\partial _{\z}\overline{\z}\partial _{\overline{\z}}\te
id+id\te \z\partial _{\z}\overline{\z}\partial
_{\overline{\z}}+\z\partial _{\z}\te \overline{\z}\partial
_{\overline{\z}}+\overline{\z}\partial _{\overline{\z}}\te \z\partial
_{\z}) .
\]
This leads to
\begin{eqnarray*}
T(f\star _{q}g) & = & e^{-h\z\partial _{\z}\overline{\z}\partial _{\overline{\z}}}\circ \mu \circ e^{h(\z\partial _{\z}\te \overline{\z}\partial _{\overline{\z}}-\overline{\z}\partial _{\overline{\z}}\te \z\partial _{\z})}(f\te g)\\
 & = & \mu \circ e^{-h(\z\partial _{\z}\overline{\z}\partial _{\overline{\z}}\te id+id\te \z\partial _{\z}\overline{\z}\partial _{\overline{\z}}+\z\partial _{\z}\te \overline{\z}\partial _{\overline{\z}}+\overline{\z}\partial _{\overline{\z}}\te \z\partial _{\z})}
  \circ e^{h(\z\partial _{\z}\te \overline{\z}\partial _{\overline{\z}}-\overline{\z}\partial _{\overline{\z}}\te \z\partial _{\z})}(f\te g)\\
 & = & \mu \circ e^{-2h(\overline{\z}\partial _{\overline{\z}}\te \z\partial _{\z})}(e^{-h\z\partial _{\z}\overline{\z}\partial_{\overline{\z}}}f\te e^{-h\z\partial _{\z}\overline{\z}\partial _{\overline{\z}}}g)\\
 & = & T(f)\star _{n}T(g)\, ,
\end{eqnarray*}
hence \( T \) is indeed an equivalence transformation from
\( \star _{n} \) to \( \star _{q} \). 
If we denote the normal ordering by \( \rho _{n} \), this new star
product can be obtained by 
$f\star _{q}g:=\rho _{q}^{-1}(\rho _{q}(f)\rho _{q}(g))$
in terms of an ``ordering prescription'' \( \rho _{q} \) given by 
\[
\rho _{q}:=\rho _{n}\circ T\, .
\]
For illustration we give the image of \( \rho _{q} \) of some simple
polynomials:
\beqa
\begin{array}{ccc}
\z^{n} & \ra  & z^{n}\\
\overline{\z}^{n} & \ra  & \overline{z}^{n}\\
%\z\overline{\z} & \ra  &  q^{-1}  z\overline{z}\;  \\
(\z\overline{\z})^n & \ra  &  q^{-n} (z\overline{z})^n. \label{rhorn}
\end{array}
\eeqa
Moreover, $\star_{q}$ is compatible with $J$:
\beq
(f\star_q g) \triangleleft J = (f\triangleleft J)\star_q g
+ f\star_q (g\triangleleft J)
\label{J-star}
\eeq
where the action of $J$ on  $\R^2$ is the obvious one.

One can easily extend the star product 
formalism to include differential forms,
which will be useful in Section \ref{sec: SW-map}.
We simply use the invertible map 
\beqa
\Omega^*     &\to& \Omega^*_q \; , \nn\\
 f = f(\z,\zbar) &\mapsto& \rho_q(f) \nn\\
\z^{-1} \zbar d\z &\mapsto&  \theta\nn\\
\z \zbar^{-1} d\zbar  &\mapsto&  \bar\theta\
\label{forms-map}
\eeqa
(extended in the obvious way)
from the differential forms on $\R^2$ to the calculus $\Omega^*_q$
defined in Section \ref{sec:calculus}, 
and define the ``star-wedge'' $\wedge_q$
on $\Omega^*$ as the pull-back of $\Omega^*_q$. 
Using the same notation 
$\theta = \z^{-1} \zbar d\z, \bar \theta =  \z \zbar^{-1} d\zbar$
as in the noncommutative algebra,  one has for example
$\theta \wedge_q \bar\theta = -q^2 \bar\theta\wedge_q\theta$
in  $\Omega^*$, 
as in $\Omega^*_q$. Clearly
$\theta \star_q f = f \star_q \theta$ in self-explanatory notation, 
and we will omit the star in this case from now on.

\subsection{$E_q(2)$-invariance of the Riemann integral}

Since there
exists an integral on the commutative space, it is natural to 
use the isomorphism $\rho$ \eq{rho} corresponding to the star product,
and define
\beq
\int^{\rho}f(z,\overline{z}):=\int \rho
^{-1}(f)(\z,\overline{\z})d\z d\overline{\z}\, .
\label{pullback-integral}
\eeq
In general, one should not expect that the integral defined in this way
is invariant under $E_q(2)$. Nevertheless, 
for the star product $\star_{q}$ defined by \( \rho _{q} \),
this integral is indeed invariant,
i.e. (\ref{eq: invariance condition for the q-integral}) is 
satisfied. We want to explain this in detail.
Consider 
\[
f(z,\overline{z})=\sum ^{\infty }_{n=-\infty }z^{n}f_{n}(z\overline{z}) \quad \in \R^2_q.
\]
Applying  $\rho ^{-1}_{q}$ gives
\[
\rho ^{-1}_{q}(f)=\sum ^{\infty }_{n=-\infty }\z^{n}\star _{q}
  \rho^{-1}_{q}(f_{n}(r^2))\, .
\] 
On the other hand, we can write the function 
\( \rho ^{-1}_{q}(f) \) in polar coordinates,
and expand it
in a Fourier series with \( r \)-dependent
coefficients
\[
\rho ^{-1}_{q}(f)=\sum ^{\infty }_{n=-\infty} e^{in\phi }\;a_{n}(r).
\]
Then
\[
a_{0}(r)=\frac{1}{2\pi }\int ^{2\pi }_{0}d\phi \, \rho^{-1}_{q}(f)(\phi ,r)\, .
\]
Since \( \z=re^{i\phi } \) and 
$\rho^{-1}_{q}(f_{n}(r^2))=f_{n}(qr^{2})$ is a function of $r^2$
by \eq{rhorn}
and using the fact \eq{J-star}
that $\star_q$ is compatible with $J$, it follows that
\[
a_{0}(r)=\rho ^{-1}_{q}(f_{0})(r^{2}) =f_{0}(qr^{2}).
\]
Therefore
\[
\int \rho^{-1}_{q}(f)(\z,\overline{\z})d\z d\overline{\z} = 2 \pi 
\int dr\, r f_{0}(qr^{2}).
\]
This agrees essentially 
with \eq{class_int}, which is indeed invariant under \( U_{q}(e(2)) \)
transformations as was shown there.

{From} now on, we will use the Riemann integral \eq{pullback-integral} 
in this context, and omit the superscript $\rho = \rho_q$ for brevity.

\subsection{Trace property and measure}

The Riemann integral does not possess the trace property, i.e.
star multiplication is not commutative under the integral. However the
trace property is necessary to obtain a gauge invariant action. We therefore look for a measure \( \mu (\z,\overline{\z}) \) such that 
\[
\int f(\z,\overline{\z})\star _{q}g(\z,\overline{\z})\, \mu
(\z,\overline{\z})d\z d\overline{\z}=\int g(\z,\overline{\z})\star
_{q}f(\z,\overline{\z})\, \mu (\z,\overline{\z})d\z d\overline{\z}\, .
\]
 Such a measure function can indeed be found.

\begin{Lemma}
\label{la: measure for trace property}

Let \( f,g \) be two arbitrary functions which vanish sufficiently fast
at infinity.
Then\begin{equation}
\label{eq: measure for integral}
\int f(\z,\overline{\z})\star _{q}g(\z,\overline{\z})\,
\frac{1}{\z\overline{\z}}d\z d\overline{\z}
=\int g(\z,\overline{\z})\star _{q}f(\z,\overline{\z})\,
\frac{1}{\z\overline{\z}}d\z d\overline{\z}
=\int f(\z,\overline{\z})g(\z,\overline{\z})\, \frac{1}{\z\overline{\z}}d\z
d\overline{\z}.
\end{equation}
\end{Lemma}

\begin{proof}

See Appendix \ref{app: proof of la about measure}.

\end{proof}
Equation (\ref{eq: measure for integral}) has also an analog on the canonical quantum plane 
$\R^2_\theta$, see e.g. \cite{Douglas:2001ba}.

A small puzzle arises here:
since the Riemannian integral is invariant under
$E_q(2)$ as we argued above, we also have the following cyclic property
\begin{equation}
\label{eq:nc-cyclic}
\int f(\z,\overline{\z})\star _{q}g(\z,\overline{\z})\star
_{q}\overline{\z}^{-1}\star _{q}\z^{-1}\, d\z d\overline{\z}
=\int g(\z,\overline{\z})\star _{q}f(\z,\overline{\z})\star
_{q}\overline{\z}^{-1}\star _{q}\z^{-1}\, d\z d\overline{\z}
\end{equation}
because of Lemma \ref{la: cyclic property of q-integral}.
These two cyclic properties are in fact equivalent, because
\begin{equation}
\label{eq: integral with q-measure prop to integral with measure}
\int G(\z,\overline{\z})\star _{q}
   (\overline{\z}^{-1}\star_{q}\z^{-1})\, d\z d\overline{\z}
=q^{-1}\int G(\z,\overline{\z})\,
\frac{1}{\z\overline{\z}}d\z d\overline{\z}\, .
\end{equation}
To see this, note that
the second equality in \eq{eq: measure for integral} implies
\beq
\int G(\z,\overline{\z})\star
_{q}\overline{\z}^{-1}\star _{q}\z^{-1}\, d\z d\overline{\z}
 = \int ((G(\z,\overline{\z})\star _{q}\overline{\z}^{-1}\star
_{q}\z^{-1})\star _{q}\z\overline{\z})\, \frac{1}{\z\overline{\z}}d\z
d\overline{\z}. 
\eeq
With \( \z\overline{\z}=q^{-1}\z\star _{q}\overline{\z} \)
which is easy to verify, it follows that
\beq
\int G(\z,\overline{\z})\star
_{q}\overline{\z}^{-1}\star _{q}\z^{-1}\, d\z d\overline{\z} 
= q^{-1}\int G(\z,\overline{\z})\,
\frac{1}{\z\overline{\z}}d\z d\overline{\z} 
\eeq
using the associativity of the star product. This shows the
equivalence of the cyclic properties \eq{eq: measure for integral} 
and \eq{eq:nc-cyclic}.

\subsection{Seiberg-Witten map} \label{sec: SW-map}

The map \( \rho _{q} \) defines a one-to-one 
correspondence between noncommutative and commutative functions, 
and we can identify \( f \) with \( \rho ^{-1}_{q}(f) \). 
We construct a Seiberg-Witten map for the noncommutative fields expressing
them by their commutative counterparts \cite{Seiberg:1999vs}: \begin{eqnarray*}
\Lambda & = & \Lambda _{\alpha }[a_{i}]\\
A_{i} & = & A_{i}[a_{i}]\\
\Psi  & = & \Psi [\psi ,a_{i}]\, .
\end{eqnarray*}
Here $a_i$ is the classical gauge field, $\a$ the classical gauge 
parameter and $\psi$ a classical matter field.
The noncommutative gauge transformations are defined as in Section 
\ref{sec:gauge-trafos} and will be spelled out below.
We assume that it is possible to expand in orders of \( h \) \begin{eqnarray}
\Lambda _{\alpha }[a_{i}] & = & \alpha +h\Lambda ^{1}_{\alpha }[a_{i}]+h^{2}\Lambda ^{2}_{\alpha }[a_{i}]+\dots \nonumber \\
A_{i}[a_{i}] & = & A^{0}_{i}+hA^{1}_{i}[a_{i}]+h^{2}A^{2}_{i}[a_{i}]+\dots \label{eq: SW expansion in h} \\
\Psi [\psi ,a_{i}] & = & \psi +h\Psi ^{1}[\psi ,a_{i}]+h^{2}\Psi ^{2}[\psi ,a_{i}]+\dots \, \, \, .\nonumber 
\end{eqnarray}
The explicit dependence on the commutative fields can be obtained 
by requiring the following \emph{consistency condition } \cite{Madore:2000en}
\begin{equation}
\label{eq:consistency}
\begin{array}{cccc}
 & (\delta _{\alpha }\delta _{\beta }-\delta _{\beta }\delta _{\alpha })\Psi  & = & \delta _{-i[\alpha ,\beta ]}\Psi \\
\Leftrightarrow  & i\delta _{\alpha }\Lambda _{\beta }-i\delta _{\beta }\Lambda _{\alpha }+[\Lambda _{\alpha }\stackrel{\star _{q}}{,}\Lambda _{\beta }] & = & i\Lambda _{-i[\alpha ,\beta ]},
\end{array}
\end{equation}
which amounts to requiring that the noncommutative gauge transformations are
induced by the commutative gauge transformations of the commutative
fields: 
\begin{eqnarray}
%\Lambda _{\alpha }[a_{i}]+\delta \Lambda _{\alpha }[a_{i}] & = & \Lambda _{\alpha }[a_{i}+\delta a_{i}]\nonumber \\
A_{i}[a_{i}]+\delta_{\L} A_{i}[a_{i}] & = & A_{i}[a_{i}+\delta_{\a} a_{i}]\label{eq:SWequ} \\
\Psi [\psi ,a_{i}]+\delta_{\L} \Psi [\psi ,a_{i}] & = & \Psi [\psi +\delta_{\a} \psi ,a_{i}+\delta_{\a} a_{i}]\, .\nonumber 
\end{eqnarray}
The consistency condition has the well-known solution \cite{Jurco:2001my}
\begin{equation}
\Lambda _{\alpha }[a_{i}]=\alpha +h\frac{1}{2}\theta ^{ij}\partial _{i}\alpha a_{j}+\mathcal{O}(h^{2})\,  .
\label{SW-solution-Lambda}
\end{equation}
This solution is hermitian for real gauge parameters $\alpha$ and for 
gauge fields $a_i$ corresponding to the hermitian connection form
$a = a_\z d\z + a_\zbar d\zbar$.
%$\overline{a_\z}=a_{\overline{\z}},
%\;\overline{a_{\overline{\z}}}=a_\z$.
%$\overline{a_1}=a_2, \;\overline{a_2}=a_1$
%(we always denote $\z_{1,2} = (\z,\obar{\z})$).
As usual, this solution is not unique. Solutions to the homogeneous
part of 
the corresponding Seiberg-Witten equation may be added leading to field
redefinitions \cite{Jurco:2001rq}.

The crucial point of our approach is that we will essentially work with
1-forms and their components $A_i$ 
w.r.t the frame $\theta^i=(\theta,\overline{\theta})$, 
\begin{equation}
A = A_i \theta^i  = \theta^i  A_i =\tilde{A}_idz^i,
\end{equation}
and that we are gauging the one-form $\Theta$ as in Section
\ref{sec:gauge-trafos}. In this way we naturally
obtain a noncommutative gauge field and field strength, with the correct 
classical limit. This is not the case if
one introduces covariant coordinates to define gauge fields and field
strengths \cite{Jurco:2001my,Meyer}, because $\theta^{ij}$ is not
constant here. 
Using \( [\Theta ,f]=df=[\lambda _{i},f]\theta ^{i} \), this led
to the gauge transformation law in the  noncommutative algebra
\begin{equation}
\label{eq: gauge trafo of Ai}
\delta _{\Lambda }A_{i}=[\lambda _{i},\Lambda ]+i[\Lambda ,A_{i}]
\end{equation}
where
\[
\lambda _{z}=\frac{1}{1-q^{-2}}\overline{z}^{-1}\, \, \,
\textrm{and}\, \, \, \lambda
_{\overline{z}}=\frac{-1}{1-q^{-2}}z^{-1} .
\]
Since the commutator with $\lambda _{i}$ satisfies the usual Leibniz rule we do not have to
introduce a "vielbein" field that transforms under gauge transformations as in \cite{Schraml:2002fi}.

In order to translate the above gauge transformation law 
to the star product approach,
we simply have to apply
\( \rho ^{-1}_{q} \). This leads to
\beq 
\label{gaugetrafo_A_for_starproduct}
\delta_{\Lambda} A_{i}=[\lambda _{i}\stackrel{\star _{q}}{,}\Lambda ]+i[\Lambda
\stackrel{\star _{q}}{,}A_{i}]\,  ,
\eeq
where we note that 
\[ \rho_{q}^{-1}(z_i^{-1})=\z_i^{-1} \,. \]
Furthermore, we remark that \[
\frac{1}{1-q^{-2}}=\frac{1}{2h}(1+h+\mathcal{O}(h^{2}))\]
such that to zeroth order we have for the gauge field
\beq    
\delta _{\alpha }A^{0}_{1} =\z\overline{\z}^{-1}\partial _{\z}\alpha \, .
\eeq
An analogous calculation for \( A_2^{0} \)
leads to the solution 
\begin{equation}
A^{0}_{i}=c_{i}a_{i}\,  ,
\label{A-zeroth-order}
\end{equation}
where
\beq
c_{\z}=\z\overline{\z}^{-1}\, \, \, \textrm{and}\, \, \,
c_{\overline{\z}}=\z^{-1}\overline{\z}\,.
\eeq
This is the solution for the gauge field, written in the basis 
$(\theta,\overline{\theta})=(c^{-1}_{\z}d\z,c^{-1}_{\overline{\z}}d\overline{\z})$ of
one-forms (cp. (\ref{de: Frame theta^i})). 
To obtain the components in the
more familiar basis $(d\z,d\overline{\z})$ we have to multiply the
above 
solution by $c^{-1}_i$, and we indeed obtain the
classical gauge field $a_i$ in zeroth order:
\begin{equation}
\tilde{A}^0_i=a_i.
\label{Atilde-zerothorder}
\end{equation}  
Defining \( c_i=: \frac{1}{1-q^{-2}} l_i \), i.e. \( l _{\z}:=\overline{\z}^{-1} \) and \( l
_{\overline{\z}}:= -\z^{-1} \), we obtain
to first order the equation 
\beq
\label{SW-equation_A}
\delta _{\alpha }A_{i}^{1}=\frac{1}{2}\theta ^{kl}\partial
_{k}l_{i}\partial _{l}\Lambda ^{1}_{\alpha }-\theta ^{kl}\partial
_{k}\alpha \partial _{l}(c_{i}a_{i})+\frac{1}{2} \theta^{kl}\partial_k l_i \partial_l\alpha \, , 
\eeq
which admits the solution 
\beq
\label{SW-solution-A}
A_{i}^{1}=c_{i}(\frac{-1}{2}\theta ^{kl}a_{k}(\partial _{l}a_{i}+F^{0}_{li}))-\frac{1}{2}\theta
^{kl}a_{k}\partial _{l}(c_{i})a_{i} + c_i a_i\, 
\eeq
where
\[F^0_{ij}:=\partial_i a_j - \partial_j a_i \]
is the usual, commutative field strength. 
This solution satisfies $\overline{A^1_1}=A^1_{2},
\overline{A^1_{2}}=A^1_1$.

We now define the noncommutative field strength as in Section 
\ref{sec:gauge-trafos},
\beq
\label{Field-strength_starproduct}
F = (\lambda_{i}\star_{q}A_{j} + A_{i}\star_{q}\lambda_{j} -iA_{i}\star_{q}A_{j})\theta
^{i}\wedge_q\theta ^{j}  = f \theta \wedge_q \bar{\theta}
\eeq 
using the ``star-calculus'' defined by \eq{forms-map},
because it satisfies the correct transformation law 
\beq
\delta f =i[\L \stackrel{\star_{q}}{,} f].
\eeq
The above solution then leads to
\begin{eqnarray}
f &=& F^0_{12}+h\big \{F^0_{12}+
\theta^{12}(F^{0}_{12}F^0_{12}-a_{\z}\partial_{\overline{\z}}F^0_{12}+a_{\overline{\z}}\partial_{\z}F^0_{12})
+\partial_{\z}\theta^{12}(a_{\z}\partial_{\overline{\z}}a_{\overline{\z}}+a_{\overline{\z}}\partial_{\overline{\z}}a_{\z}+2a_{\overline{\z}}\partial_{\z}a_{\overline{\z}})
 \nonumber \\ 
& &
+\partial_{\overline{\z}}\theta^{12}(a_{\z}\partial_{\z}a_{\overline{\z}}+a_{\overline{\z}}\partial_{\z}a_{\z}+2a_{\z}\partial_{\overline{\z}}a_{\z})
\big \}+\mathcal{O}(h^2) \, .
\label{expanded_Fieldstrength}
\end{eqnarray}
We can now write down the following action using the classical integral:
\begin{equation}
S:= \frac{1}{2} \int f \star_q f \, 
\frac{1}{\z\overline{\z}} d\z d\overline{\z} \, .
\end{equation}
Recall that the measure function 
$\mu(\z,\overline{\z})=\frac{1}{\z\overline{\z}}$ 
is necessary to ensure gauge invariance of the action, using
the trace property of the integral by Lemma \ref{la: measure for trace 
property}.
This action can be written in terms of commutative fields using the
above result:
\begin{eqnarray}
S & = & \int d\z d\overline{\z} \frac{1}{\z\overline{\z}} \, \big \{\frac{1}{2}F^{0}_{12}F^0_{12}+h\big
(F^{0}_{12}F^0_{12}+\theta^{12}(F^{0}_{12}F^0_{12}F^0_{12}-a_{\z}F^{0}_{12}\partial _{\overline{\z}}F^{0}_{12}+a_{\overline{\z}}F^{0}_{12}\partial
 _{\z}F^{0}_{12}) \nonumber \\
 & & +\partial_{\overline{\z}} \theta^{12}F^{0}_{12}(2a_{\z}\partial _{\overline{\z}}a_{\z}+a_{\z}\partial
 _{\z}a_{\overline{\z}}+a_{\overline{\z}}\partial _{\z}a_{\z})  
\\
 & &+\partial_{\z} \theta^{12}F^{0}_{12}(2a_{\overline{\z}}\partial _{\z}a_{\overline{\z}}+a_{\overline{\z}}\partial
 _{\overline{\z}}a_{\z}+a_{\z}\partial
 _{\overline{\z}}a_{\overline{\z}})
\big )\big \}+\mathcal{O}(h^{2}) \, . \nonumber
\end{eqnarray}
Observe that this action is also the Seiberg-Witten form of 
\eq{action-algebra},
because 
\begin{equation}
S=\frac{1}{2} \int f \star_q f \, 
\frac{1}{\z\overline{\z}} d\z d\overline{\z} = 
\frac q2 \int f \star_q f \,\star_q (\overline{\z}^{-1}\star_q
\z^{-1}) d\z d\overline{\z}\, 
\end{equation}
using (\ref{eq: integral with q-measure prop to integral with
measure}). We see that as in the algebraic approach of Section 
\ref{sec:gauge-trafos},
gauge invariance requires a
measure function $\mu(\z,\overline{\z})=\frac{1}{\z\overline{\z}}
d\z d\overline{\z}$ which breaks translation invariance. 
However, one should realize that even without this 
measure function, this ``classical'' action would not be invariant 
under $E(2)$, because the star product is not compatible with 
the symmetry (only for rotations \eq{J-star} holds). This would only
be the case if one could find a star product on $\R^2_q$ which is 
compatible with the coproduct of $E_q(2)$, cp. 
\cite{Blohmann:2002wx,Grosse:2001pr}.

\subsection{The classical limit and the measure function}

The measure function 
$\mu(\z,\overline{\z})=\frac{1}{\z\overline{\z}} d\z d\overline{\z}$ 
survives in the classical limit $q\rightarrow 1$. 
If we want a deformation of the classical theory, 
this should not be the case. We therefore would like to get rid of
this measure function in the classical limit. This can be
achieved by multiplying 
the action with a gauge-covariant expression\footnote{
This was suggested by Peter Schupp.}, which in the
classical limit exactly cancels the measure function $\mu$. 
For this purpose we introduce covariant coordinates \cite{Madore:2000en}:
\begin{equation}
\mathcal{Z}_i:=\z_i+\mathcal{A} _i \,.
\end{equation}
Here $\mathcal{A}_i$ should not be confused with $A_i$. 
The one-form $A_i \theta^i$ is a noncommutative analog of the 
classical gauge field, % $a_i d\z^i$. This is 
because its gauge transformation law (\ref{gaugetrafo_A_for_starproduct}) 
is the noncommutative generalization of the classical gauge
transformation law. Indeed, 
we recovered the classical gauge field $a_i$ with respect to the basis
$d\z,d\overline{\z}$ (\ref{Atilde-zerothorder}) 
in zeroth order of $h$. In contrast, 
the covariant coordinates are used here just as a quantity which 
transforms covariantly and reduces to the usual coordinates
in the classical limit, in order to cancel the measure 
function. We will see that $\mathcal{A}_i$
does not reduce to the classical gauge field for $q \to 1$. 
Requiring the covariant transformation rule $\delta \mathcal{Z}_i=i[\Lambda\stackrel{\star_q}{,}\mathcal{Z}_i]$ leads to the following gauge
transformation rule for $\mathcal{A}_i$
\begin{equation}
\delta \mathcal{A}_i=i[\z_i\stackrel{\star_q}{,}\Lambda]+i[\Lambda\stackrel{\star_q}{,}\mathcal{A}_i] \,.
\end{equation} 
As  before we can express $\mathcal{A}_i$ in terms of commutative fields
by solving the corresponding
Seiberg-Witten equations. This gives \cite{Jurco:2001my}
\begin{equation}
\mathcal{A}^i=h\theta^{ij}a_j+h^2\frac{1}{2}\theta^{kl}a_l(\partial_k(\theta^{ij}a_j)-\theta^{ij}F^{0}_{jk})+\mathcal{O}(h^3)
\,.
\end{equation}
In principle, covariant coordinates may be used to define noncommutative gauge
fields and 
covariant expressions such as field strength  \cite{Madore:2000en,Meyer}. However,
the above equation shows that 
gauge fields and field strengths defined in that way 
do not lead to the classical gauge field $a_i$ and field strength
$F^0_{ij}$ in the limit $h\rightarrow0$ whenever the
Poisson-structure is not constant and not invertible, 
as is the case here\footnote{To obtain in the classical limit the
classical gauge field $a_i$ we have to invert $\theta^{ij}$ and write
$\frac{1}{h}\theta^{-1}_{ij}\mathcal{A}^j$. This is only defined if
$\theta$ is 
invertible, and even then it spoils the covariant
transformation property whenever $\theta$ is not constant. 
To maintain covariance one has to "invert $\theta$ covariantly" as
done in \cite{Meyer}, leading to complicated expressions. 
The approach that we propose in Section \ref{sec: SW-map} does not have these problems. Gauging
the one-form $\Theta$ instead of the coordinates leads very naturally 
to a noncommutative gauge-field (\ref{gaugetrafo_A_for_starproduct}) and
field strength (\ref{Field-strength_starproduct}). Compare also with
\cite{Schraml:2002fi}, 
where a different approach using a "vielbein" is discussed.}.
Nevertheless they are a convenient tool for our purpose, because
they satisfy 
\begin{equation}
\mathcal{Z}\star_q\overline{\mathcal{Z}}\rightarrow \z\overline{\z}
\end{equation}
for $q\rightarrow 1$, and
\begin{equation}
\delta (\mathcal{Z}\star_q\overline{\mathcal{Z}}) = i[\Lambda\stackrel{\star_q}{,}\mathcal{Z}\star_q\overline{\mathcal{Z}}] \,.
\end{equation}
Now we can define a gauge-invariant action with the correct 
classical limit:
\begin{equation}
S^{'}:=\frac{1}{2} \int f \star_q f \star_q \mathcal{Z}\star_q\overline{\mathcal{Z}} \, \frac{1}{\z\overline{\z}} d\z d\overline{\z}\,.
\end{equation}
Expanded up to first order of $h$ we obtain
\begin{eqnarray}
S^{'} &=&\int d\z d\overline{\z} \, \frac{1}{2}F^0_{12}F^0_{12}+h\big (F^0_{12}F^0_{12}+\theta^{12}(F^{0}_{12}F^0_{12}F^0_{12}-a_{\z}F^{0}_{12}\partial _{\overline{\z}}F^{0}_{12}+a_{\overline{\z}}F^{0}_{12}\partial
 _{\z}F^{0}_{12}) \nonumber \\
 & & +\partial_{\overline{\z}} \theta^{12}F^{0}_{12}(2a_{\z}\partial _{\overline{\z}}a_{\z}+a_{\z}\partial
 _{\z}a_{\overline{\z}}+a_{\overline{\z}}\partial _{\z}a_{\z})  \label{expanded_action} \\
 & & +\partial_{\z} \theta^{12}F^{0}_{12}(2a_{\overline{\z}}\partial _{\z}a_{\overline{\z}}+a_{\overline{\z}}\partial
 _{\overline{\z}}a_{\z}+a_{\z}\partial
 _{\overline{\z}}a_{\overline{\z}})
%-3F^{0^2}_{12}
\nonumber \\
 & & +\frac{1}{\z\overline{\z}}\theta^{12}F^{0}_{12}F^0_{12}(\overline{\z}a_{\overline{\z}}-\z a_{\z})
 -F^{0}_{12}F^0_{12}+\z\partial_{\z}(F^{0}_{12}F^0_{12})-\overline{\z} \partial_{\overline{\z}}(F^{0}_{12}F^0_{12}) \big )+\mathcal{O}(h^{2}) \, . \nonumber
\end{eqnarray}
This reduces indeed to a Yang-Mills theory  in the classical limit.
However, choosing  $\mathcal{Z}\star_q\overline{\mathcal{Z}}$ is only one
possibility to cancel $\frac{1}{\z\overline{\z}}$.
There are other expressions which are gauge-covariant, and lead
to the same classical limit. Our choice is motivated by simplicity.

\section*{Acknowledgements}

We want to thank in particular Peter Schupp 
for his collaboration and many useful suggestions in the initial stage 
of this project. We also wish to thank P. Aschieri, W. Behr, F. Bachmaier,
A. Sykora and J. Wess for useful discussions.

\appendix

\section{Mathematical Appendix\label{appendix: mathematical}}

\subsection{Coaction and action}

\begin{Def}\label{de: coaction of a Hopfalgebra on a algebra}

A left coaction of a Hopf algebra $\mathcal{H}$ on an algebra
$\mathcal{A}$ is a linear mapping 
\be
\rho : \mathcal{A} \longrightarrow \mathcal{H}\otimes \mathcal{A}
\ee
which satisfies 
\be
\begin{array}{lc}
(\mathrm{id} \otimes \rho)\circ \rho = (\Delta \otimes
\mathrm{id})\circ \rho, 
    \qquad (\varepsilon \otimes \mathrm{id})\circ \rho = \mathrm{id} \\
\rho(ab)=\rho(a)\rho(b),\qquad 
\rho(1)=1\otimes1. \qquad \qquad 
\end{array} \ee In Sweedler notation, one writes %\cite[p. 32]{Klimyk:1997eb} 
\[
\rho (a)=:a_{(-1)}\te a_{(0)}\,  .\]
\( \mathcal{A} \)  is then called a left \( \mathcal{H} \)-comodule algebra.\\
\end{Def}

\begin{Def}\label{de: def of action on an algebra}

\textit{A Hopf algebra \( \mathcal{H} \) is acting on
an algebra \( \mathcal{A} \) from the right  if \( \mathcal{A} \)
if there is an action $\lt : \mathcal{A}\tens\mathcal{H}\to \mathcal{A}$
which satisfies
%is a right \( \mathcal{H} \)-module such that 
%\( m:\mathcal{A}\ten \mathcal{A}\ar \mathcal{A} \)
%and \( \eta :\mathbb {C}\ar \mathcal{A} \) are right  \( \mathcal{H} \)-module
%homomorphisms, that means if 
}
\textit{\emph{\begin{equation}
\label{action}
ab\lt h=(a\te b)\lt \Delta (h)=(a\lt h_{(1)})(b\lt h_{(2)})\, \, \, \, \, \, \textrm{and}\, \, \, \, \, \, 1\lt h=\varepsilon (h)1
\end{equation}
}} \textit{for any \( h\in \mathcal{H} \) and \( a,b\in \mathcal{A} \).
\( \mathcal{A} \)  is then called a right \( \mathcal{H}- \)module
algebra.}

\end{Def}

By (\ref{de: action from coaction}), these two notions are dual to
each other. 
There are obvious analogs
replacing left with right everywhere.

For the action of \( J,T \) and \( \overline{T} \) on
the generators \( z,\overline{z} \), we obtain
\beqa
\label{eq: action of Uq on generators z, zbar}
z\lt T & = & 1, \qquad
z\lt \overline{T} =  0, \qquad\quad z\lt J  =  iz\nn\\
\overline{z}\lt T & = &0 , \qquad 
\overline{z}\lt \overline{T}  =  -q^{2}, \qquad
\overline{z}\lt J  =  -i\overline{z}\; . \nn
\eeqa

The action on arbitrary functions is calculated in the following subsection.

\subsection{The right action of \protect\( U_{q}(e(2))\protect \) on \protect\( \mathbb {R}_{q}^{2}\protect \)\label{appendix: Action of Uq on A_nc}}

Knowing the structure maps (\ref{eq: commrel and structure maps for Uq(e(2))})
for \( J,T,\overline{T}\in U_{q}(e(2)) \) and their
action on \( z,\overline{z} \) given above,
%(see (\ref{eq: action of Uq on generators z, zbar}))
we can determine the action of \( J,T,\overline{T} \) on arbitrary
functions using \( (xy)\lt U=(x\lt U_{(1)})(y\lt U_{(2)}) \) for
arbitrary \( x,y\in \mathbb {R}_{q}^{2},\, U\in U_{q}(e(2)) \). 
Since an arbitrary
function \( f(z,\overline{z})\in \mathbb {R}_{q}^{2} \) can be written
as \( f(z,\overline{z})=\sum _{k\in \mathbb
  {Z}}z^{k}f_{k}(z\overline{z}) \), 
it is sufficient to know the action on the terms
\[
z^{k}f(z\overline{z})\, ,
\]
where \( f \) is a formal power series in \( z\overline{z} \).
We will derive the formulas even for negative powers of \( z\overline{z} \),
i.e. \( f(z\overline{z})=\sum _{l\in \mathbb {Z}}a_{l}(z\overline{z})^{l} \).
We start with the action on \( z^{k} \):

\begin{Cl}

For \( k\in \mathbb {Z} \) we have\begin{eqnarray}
z^{k}\lt T & = & \frac{1-q^{-2k}}{1-q^{-2}}z^{k-1}\nonumber \\
z^{k}\lt \overline{T} & = & 0\label{app: action of Uq on z^k} \\
z^{k}\lt J & = & i^{k}z^{k}\, \, \, .\nonumber 
\end{eqnarray}

\end{Cl}

\begin{proof}

The first equation can be shown by induction,
using \( z\lt T=1 \) and $z^{-1}\lt T=-q^{2}z^{-2}$,
which follows from
\[
0=1\lt T=(z^{-1}z)\lt T=(z^{-1}\lt T)(z\lt q^{2iJ})+z^{-1}(z\lt
T)=(z^{-1}\lt T)q^{-2}z+z^{-1}.
\]
The last two equations finally follow immediately with \( z\lt \overline{T}=0 \),
\( z\lt J=iz \) and \( \Delta (\overline{T})=\overline{T}\te
q^{2iJ}+1\te \overline{T}. 
\)
%first for \( k>0 \) and then with \( 1\lt \overline{T}=0=z^{-1}\lt \overline{T} \)
%also for \( k\leq 0 \). 

\end{proof}

The action on 
\( f(z\overline{z})=\sum _{l\in \mathbb {Z}}a_{l}(z\overline{z})^{l}
\)
follows from
%we start considering the action on a summand \( (z\overline{z})^{l} \):
\begin{Cl}

For \( l\in \mathbb {Z} \) we have\begin{eqnarray}
(z\overline{z})^{l}\lt T & = & q^{2}\frac{1-q^{-2l}}{1-q^{-2}}(z\overline{z})^{l-1}\overline{z}\nonumber \\
(z\overline{z})^{l}\lt \overline{T} & = & -q^{2}\frac{1-q^{2l}}{1-q^{2}}(z\overline{z})^{l-1}z\label{app: action on (zzbar)^l} \\
(z\overline{z})^{l}\lt J & = & (z\overline{z})^{l}\, \, \, .\nonumber 
\end{eqnarray}

\end{Cl}

\begin{proof}

The last equation follows immediately with 
\( z\lt J=iz \), \( \overline{z}\lt J=-i\overline{z} \).
%and \( \Delta (J)=J\te 1+1\te J \). 
The first equation follows again
by induction, starting with
%\begin{itemize}
%\item We have 
\( (z\overline{z})\lt T=(z\lt T)(\overline{z}\lt
q^{2iJ})+z(\overline{z}\lt T)=q^{2}\overline{z} \),
and concluding inductively
\begin{eqnarray*}
(z\overline{z})^{l+1}\lt T & = & ((z\overline{z})^{l}\lt T)((z\overline{z})\lt q^{2iJ})+(z\overline{z})^{l}((z\overline{z})\lt T)\\
 & = & q^{2}\frac{1-q^{-2l}}{1-q^{-2}}(z\overline{z})^{l-1}\overline{z}(z\overline{z})+(z\overline{z})^{l}q^{2}\overline{z}\\
 & = & q^{2}\frac{1-q^{-2l-2}}{1-q^{-2}}(z\overline{z})^{l}\overline{z}
\end{eqnarray*}
for \( l >0. \)
%\end{itemize}
If \( l=0 \), then \( 1\lt T=0 \), which is consistent with the
claim. To derive the action of \( T \) on \( (z\overline{z})^{-1} \)
we calculate\[
0=((z\overline{z})^{-1}(z\overline{z}))\lt T=((z\overline{z})^{-1}\lt T)z\overline{z}+(z\overline{z})^{-1}((z\overline{z})\lt T)=((z\overline{z})^{-1}\lt T)z\overline{z}+(z\overline{z})^{-1}q^{2}\overline{z}\]
hence
\[
(z\overline{z})^{-1}\lt T=-(z\overline{z})^{-2}\overline{z},\]
consistent with (\ref{app: action on (zzbar)^l}). For
\( l<0 \) the claim follows similarly by induction, and
the second equation follows also inductively.

\end{proof}

Putting these results together and using \( f(z\overline{z})=\sum _{l\in \mathbb {Z}}a_{l}(z\overline{z})^{l} \)
we obtain\begin{eqnarray*}
z^{k}f(z\overline{z})\lt T & = & (z^{k}\lt T)(f(z\overline{z})\lt q^{2iJ})+z^{k}(f(z\overline{z})\lt T)\\
% & = & \frac{1-q^{-2k}}{1-q^{-2}}z^{k-1}f(z\overline{z})+z^{k}\sum _{l\in \mathbb {Z}}a_{l}q^{2}\frac{1-q^{-2l}}{1-q^{-2}}(z\overline{z})^{l-1}\overline{z}\\
 & = & \frac{1-q^{-2k}}{1-q^{-2}}z^{k-1}f(z\overline{z})+z^{k-1}\sum _{l\in \mathbb {Z}}a_{l}q^{2}\frac{1-q^{-2l}}{1-q^{-2}}q^{2(l-1)}(z\overline{z})^{l}\\
% & = & \frac{z^{k-1}}{1-q^{-2}}((1-q^{-2k})f(z\overline{z})+\sum _{l\in \mathbb {Z}}a_{l}(q^{2l}-1)(z\overline{z})^{l})\\
 & = & \frac{z^{k-1}}{1-q^{-2}}(f(q^{2}z\overline{z})-q^{-2k}f(z\overline{z}))\, .
\end{eqnarray*}
A similar calculation finally leads to (\ref{eq: action of J,T,Tbar on z^kf(zzbar)}).

\subsection{Proof of Lemma \ref{la: Theta generates d}}
\label{app:proof-La2}

\begin{proof}

Since the \( \theta ^{i} \) commute with all functions, we have
\[
[\Theta ,f]=\theta
[\frac{1}{1-q^{-2}}\overline{z}^{-1},f]-\overline{\theta
}[\frac{1}{1-q^{-2}}z^{-1},f]\, .
\]
 Plugging in the explicit expressions (\ref{de: Frame theta^i}) for
\( \theta ^{i} \) we find
\begin{eqnarray*}
[\Theta ,f] %& = & z^{-1}\overline{z}dz[\frac{1}{1-q^{-2}}\overline{z}^{-1},f]-d\overline{z}z\overline{z}^{-1}[\frac{1}{1-q^{-2}}z^{-1},f]\\
 & = & dzz^{-1}\overline{z}[\frac{1}{1-q^{-2}}\overline{z}^{-1},f]-d\overline{z}z\overline{z}^{-1}[\frac{1}{1-q^{-2}}z^{-1},f] \, ,
\end{eqnarray*}
using the commutation relations (\ref{eq: comm-rel dz,dzbar with z,zbar}).
Taking \( f=z \) and \( f=\overline{z} \) we get 
%with the commutation relations of the space coordinates 
\begin{equation}
\label{pr: deriv on generator z}
z^{-1}\overline{z}[\frac{1}{1-q^{-2}}\overline{z}^{-1},z]=\frac{1}{1-q^{-2}}-\frac{q^{-2}}{1-q^{-2}}=1
\end{equation}
 and
\begin{equation}
\label{pr: deriv on generator zbar}
z\overline{z}^{-1}[\frac{1}{1-q^{-2}}z^{-1},z]=0\, .
\end{equation}
Thus \( [\Theta ,z]=dz \), and similarly 
\( [\Theta ,\overline{z}]=d\overline{z} \).
Hence the claim is
true on the generators of the algebra of functions, and 
since \( [\Theta , .] \) is a derivation we can  conclude
that 
\[
df=[\Theta ,f]
\]
for all functions \( f \). 

To show \( d\Theta =\Theta ^{2}=0 \), consider
\begin{eqnarray*}
((1-q^{-2})\Theta )^{2} & = & (\theta \overline{z}^{-1}-\overline{\theta }z^{-1})^{2}=(q^{-2}z^{-1}dz-\overline{z}^{-1}d\overline{z})^{2}\\
 & = & -q^{-2}z^{-1}dz\overline{z}^{-1}d\overline{z}-\overline{z}^{-1}d\overline{z}q^{-2}z^{-1}dz\\
 & = & -q^{-4}z^{-1}\overline{z}^{-1}dzd\overline{z}-\overline{z}^{-1}z^{-1}d\overline{z}dz=0\,  ,
\end{eqnarray*}
using the commutation relations (\ref{coac}), 
(\ref{eq: comm-rel dz,dzbar with z,zbar})
and (\ref{eq: comm-rel for differentials}). Furthermore,
\[
(1-q^{-2})d\Theta
=d(q^{-2}z^{-1}dz-\overline{z}^{-1}d\overline{z})=-q^{-4}z^{-2}dzdz+q^{2}\overline{z}^{-2}d\overline{z}d\overline{z}=0
\]
 where we used \( d(z^{-1})=-q^{-2}z^{-2}dz \) and \( d(\overline{z}^{-1})=-q^{2}\overline{z}^{-2}d\overline{z} \),
which follows from
the \( q- \)Leibniz rule applied to
\( 0=d1=d(zz^{-1})=d(\bar{z}\bar{z}^{-1}). \)
\end{proof} \textbf{}

\subsection{Proof of Lemma \ref{la: measure for trace property}\label{app: proof of la about measure}}

\begin{proof}

We have
\begin{eqnarray*}
\int \frac{d\z d\overline{\z}}{\z\overline{\z}}\, f\star _{q}g=\int
\frac{d\z d\overline{\z}}{\z\overline{\z}}\, fg+\int \frac{d\z d\overline{\z}}{\z\overline{\z}}\, \mu \circ \sum ^{\infty }_{n=1}\frac{h^{n}}{n!}(\sum ^{2}_{i_{1},j_{1}=1}\varepsilon ^{i_{1}j_{1}}\z^{i_{1}}\frac{\partial }{\partial \z^{i_{1}}}\te \z^{j_{1}}\frac{\partial }{\partial \z^{j_{1}}}) & \\
(\sum ^{2}_{i_{2},j_{2}=1}\varepsilon ^{i_{2}j_{2}}\z^{i_{2}}\frac{\partial }{\partial \z^{i_{2}}}\te \z^{j_{2}}\frac{\partial }{\partial \z^{j_{2}}})\dots (\sum ^{2}_{i_{n},j_{n}=1}\varepsilon ^{i_{n}j_{n}}\z^{i_{n}}\frac{\partial }{\partial \z^{i_{n}}}\te \z^{j_{n}}\frac{\partial }{\partial \z^{j_{n}}})(f\te g)\, . & 
\end{eqnarray*}
Consider the \( n \) -th term of the sum on the right hand
side:
\begin{eqnarray*}
\int \frac{d\z d\overline{\z}}{\z\overline{\z}}\, \frac{h^{n}}{n!}\mu \circ (\sum ^{2}_{i_{1},j_{1}=1}\varepsilon ^{i_{1}j_{1}}\z^{i_{1}}\frac{\partial }{\partial \z^{i_{1}}}\te \z^{j_{1}}\frac{\partial }{\partial \z^{j_{1}}})(\sum ^{2}_{i_{2},j_{2}=1}\varepsilon ^{i_{2}j_{2}}\z^{i_{2}}\frac{\partial }{\partial \z^{i_{2}}}\te \z^{j_{2}}\frac{\partial }{\partial \z^{j_{2}}}) &  & \\
\dots (\sum ^{2}_{i_{n},j_{n}=1}\varepsilon ^{i_{n}j_{n}}\z^{i_{n}}\frac{\partial }{\partial \z^{i_{n}}}\te \z^{j_{n}}\frac{\partial }{\partial \z^{j_{n}}})(f\te g)\, . &  & 
\end{eqnarray*}
Introducing the short hand notation 
\[
f^{'}\te g^{'}:=(\sum ^{2}_{i_{2},j_{2}=1}\varepsilon
^{i_{2}j_{2}}\z^{i_{2}}\frac{\partial }{\partial \z^{i_{2}}}\te
\z^{j_{2}}\frac{\partial }{\partial \z^{j_{2}}})\dots (\sum
^{2}_{i_{n},j_{n}=1}\varepsilon ^{i_{n}j_{n}}\z^{i_{n}}\frac{\partial
}{\partial \z^{i_{n}}}\te \z^{j_{n}}\frac{\partial }{\partial
  \z^{j_{n}}})(f\te g),
\]
the \( n- \)th term of the sum can be written as
\[
\int \frac{d\z d\overline{\z}}{\z\overline{\z}}\, \frac{h^{n}}{n!}\mu \circ (\sum ^{2}_{i_{1},j_{1}=1}\varepsilon ^{i_{1}j_{1}}\z^{i_{1}}\frac{\partial }{\partial \z^{i_{1}}}\te \z^{j_{1}}\frac{\partial }{\partial \z^{j_{1}}})(f^{'}\te g^{'})\]
\[
=\frac{h^{n}}{n!}\int d\z d\overline{\z}\, \sum
^{2}_{i_{1},j_{1}=1}\varepsilon ^{i_{1}j_{1}}\frac{\partial }{\partial
  \z^{i_{1}}}(f^{'})\frac{\partial }{\partial \z^{j_{1}}}(g^{'})\,  .
\]
For \( n>0 \), this leads after partial integration (assuming 
that the functions vanish at infinity) to 
\[
-\frac{h^{n}}{n!}\int d\z d\overline{\z}\, \sum
^{2}_{i_{1},j_{1}=1}\varepsilon ^{i_{1}j_{1}}f^{'}\frac{\partial
}{\partial \z^{i_{1}}}\frac{\partial }{\partial \z^{j_{1}}}(g^{'})=0\,
.
\]
 This is valid for any summand corresponding to \( n>0 \), so that only
the zeroth order term does not vanish. Hence we find indeed
\[
\int \frac{d\z d\overline{\z}}{\z\overline{\z}}\, f\star _{q}g=\int
\frac{d\z d\overline{\z}}{\z\overline{\z}}\, fg\, .\]

\end{proof}

\bibliographystyle{diss}
\bibliography{literature}

\begin{thebibliography}{10}

\bibitem{Douglas:2001ba}
M.~R. Douglas and N.~A. Nekrasov,
\newblock {\em Noncommutative field theory},
\newblock Rev. Mod. Phys. {\bf 73}, 977 (2001), hep-th/0106048.

\bibitem{Krajewski:1999ja}
T.~Krajewski and R.~Wulkenhaar,
\newblock {\em Perturbative quantum gauge fields on the noncommutative torus},
\newblock Int. J. Mod. Phys. {\bf A15}, 1011 (2000), hep-th/9903187.

\bibitem{Connes:1998cr}
A.~Connes, M.~R. Douglas, and A.~Schwarz,
\newblock {\em Noncommutative geometry and matrix theory: Compactification on
  tori},
\newblock JHEP {\bf 02}, 003 (1998), hep-th/9711162.

\bibitem{Seiberg:1999vs}
N.~Seiberg and E.~Witten,
\newblock {\em String theory and noncommutative geometry},
\newblock JHEP {\bf 09}, 032 (1999), hep-th/9908142.

\bibitem{Calmet:2001na}
X.~Calmet, B.~Jurco, P.~Schupp, J.~Wess, and M.~Wohlgenannt,
\newblock {\em The standard model on non-commutative space-time},
\newblock Eur. Phys. J. {\bf C23}, 363 (2002), hep-ph/0111115.

\bibitem{Pawelczyk:2003nb}
J.~Pawelczyk and H.~Steinacker,
\newblock {\em Algebraic brane dynamics on SU(2): Excitation spectra},
\newblock (2003), hep-th/0305226.

\bibitem{Chaichian:1999wy}
M.~Chaichian, A.~Demichev, and P.~Presnajder,
\newblock {\em Quantum field theory on the noncommutative plane with {$E_q(2)$}
  symmetry},
\newblock J. Math. Phys. {\bf 41}, 1647 (2000), hep-th/9904132.

\bibitem{Schraml:2002fi}
S.~Schraml,
\newblock {\em Non-Abelian gauge theory on q-quantum spaces},
\newblock (2002), hep-th/0208173.

\bibitem{Schupp:1992ex}
P.~Schupp, P.~Watts, and B.~Zumino,
\newblock {\em The Two-dimensional quantum Euclidean algebra},
\newblock Lett. Math. Phys. {\bf 24}, 141 (1992), hep-th/9206024.

\bibitem{Wess:1991vh}
J.~Wess and B.~Zumino,
\newblock {\em Covariant Differential Calculus on the Quantum Hyperplane},
\newblock Nucl. Phys. Proc. Suppl. {\bf 18B}, 302 (1991).

\bibitem{Chaichian:1994ft}
M.~Chaichian and A.~P. Demichev,
\newblock {\em q deformed path integral},
\newblock Phys. Lett. {\bf B320}, 273 (1994), hep-th/9310001.

\bibitem{Madore:Buch}
J.~Madore,
\newblock {\em An Introduction to Noncommutative Differential Geometry and its
  Physical Applications},
\newblock Cambridge University Press, secon ed. (1999) 200 p. (London
  Mathematical Society Lecture Note Series 257).

\bibitem{Cerchiai:2000qu}
B.~L. Cerchiai, G.~Fiore, and J.~Madore,
\newblock {\em Geometrical Tools for Quantum Euclidean Spaces},
\newblock Commun. Math. Phys. {\bf 217}, 521 (2001), math.qa/0002007.

\bibitem{Koe}
H.~T. Koelink,
\newblock {\em {The quantum group of plane motions and the Hahn-Exton
  $q$-Bessel function}},
\newblock Duke Math. J. {\bf 76}, 483 (1994).

\bibitem{Steinacker:1995jh}
H.~Steinacker,
\newblock {\em Integration on quantum Euclidean space and sphere in N-
  dimensions},
\newblock J. Math. Phys. {\bf 37 Nr.9} (1996), q-alg/9506020.

\bibitem{Madore:2000en}
J.~Madore, S.~Schraml, P.~Schupp, and J.~Wess,
\newblock {\em Gauge theory on noncommutative spaces},
\newblock Eur. Phys. J. {\bf C16}, 161 (2000), hep-th/0001203.

\bibitem{Grosse:2000Fuzzy1}
H.~Grosse, J.~Madore, and H.~Steinacker,
\newblock {\em Field theory on the q-deformed fuzzy sphere. I},
\newblock J. Geom. Phys. {\bf 38}, 308 (2001), hep-th/0005273.

\bibitem{Meyer}
F.~Meyer,
\newblock {\em Models of Gauge Field Theory on Noncommutative Spaces},
\newblock Diploma-Thesis, {LMU-M\"unchen}, Chair Prof. J. Wess  (2003),
  hep-th/0308186.

\bibitem{Grosse:2001pr}
H.~Grosse, J.~Madore, and H.~Steinacker,
\newblock {\em Field theory on the q-deformed fuzzy sphere. II: Quantization},
\newblock J. Geom. Phys. {\bf 43}, 205 (2002), hep-th/0103164.

\bibitem{Bayen:1978ha}
F.~Bayen, M.~Flato, C.~Fronsdal, A.~Lichnerowicz, and D.~Sternheimer,
\newblock {\em Deformation Theory and Quantization},
\newblock Ann. Phys. {\bf 111}, 61 (1978).

\bibitem{Jurco:2001my}
B.~Jurco, P.~Schupp, and J.~Wess,
\newblock {\em Nonabelian noncommutative gauge theory via noncommutative extra
  dimensions},
\newblock Nucl. Phys. {\bf B604}, 148 (2001), hep-th/0102129.

\bibitem{Jurco:2001rq}
B.~Jurco, L.~{M\"oller}, S.~Schraml, P.~Schupp, and J.~Wess,
\newblock {\em Construction of non-Abelian gauge theories on noncommutative
  spaces},
\newblock Eur. Phys. J. {\bf C21}, 383 (2001), hep-th/0104153.

\bibitem{Blohmann:2002wx}
C.~Blohmann,
\newblock {\em Covariant realization of quantum spaces as star products by
  Drinfeld twists},
\newblock J. Math. Phys. {\bf 44}, 4736 (2003), math.qa/0209180.

\end{thebibliography}

\end{document}